\documentclass[
 reprint,
 amsmath,amssymb,
 aps,
 pra,
]{revtex4-2}

\usepackage{amsmath}
\usepackage{amsthm}
\usepackage{amsfonts}
\usepackage{graphicx}
\usepackage{dcolumn}
\usepackage{bm}
\usepackage[colorlinks,urlcolor=blue,linkcolor=blue,citecolor=blue]{hyperref}
\usepackage[capitalise]{cleveref}
\usepackage{tabularray}
\usepackage{xspace}

\makeatletter
\AddToHook{cmd/appendix/before}{
  \crefalias{section}{appendix}
}
\makeatother

\newcommand{\intd}{\,\mathrm{d}}
\newcommand{\T}{^\intercal} 
\newcommand{\invec}[2]{\left(#1, #2 \right)\T}
\newcommand{\inmat}[4]{\left(\begin{smallmatrix}#1 & #2 \\ #3 & #4 \end{smallmatrix}\right)}
\newcommand{\dvec}[2]{\begin{pmatrix} #1 \\ #2 \end{pmatrix}}

\newcommand{\ai}{atom interferometer\xspace} 
\newcommand{\ais}{atom interferometers\xspace} 
\newcommand{\mom}{momentum\xspace} 
\newcommand{\moma}{momenta\xspace} 
\newcommand{\lcp}{LC$^+$\xspace}
\newcommand{\lcm}{LC$^-$\xspace}
\newcommand{\lcpm}{LC$^{\pm}$\xspace}
\newcommand{\figwidth}{1.0\linewidth}

\begin{document}

\title{Exact Semiclassical Phase Shifts for Relativistic Atom Interferometers in Flat Spacetime}

\author{Hunter Swan}
 \email{orswan@stanford.edu}
\author{Jason M. Hogan}
 \email{hogan@stanford.edu}
\affiliation{
 Department of Physics, Stanford University, Stanford, California 94305, USA
}
\date{\today}

\begin{abstract}
Atom interferometry is a sensitive tool for measuring relativistic effects, but there are no known non-trivial exact solutions for relativistic atom interferometer phase shifts.  Here we derive relativistically exact expressions within the usual semiclassical approximation for a wide range of experimentally interesting atom interferometer pulse sequences in flat spacetime, including Mach-Zehnder, resonant, and large momentum transfer interferometer geometries.  As an example, the leading order phase shift $\omega_a g T^2/c$ for a Mach-Zehnder clock atom interferometer is found to become $\omega_a \left(1 + \frac{\omega_a}{2m}\right)(e^{-gT/c}-1)^2 c/g$ when all relativistic kinematics are included.  We calculate exact phase shifts for both clock (single-photon) interferometers and Raman or Bragg (two-photon) interferometers. 
\end{abstract}

\maketitle

\section{Introduction}
From the first demonstration of light-pulse atom interferometry in the Kasevich-Chu experiment \cite{kasevich1992measurement}, it has been recognized that this phenomenon has a deep connection to gravity, and by extension to relativity.  Many proposals have been made for using atom interferometers to measure relativistic phenomena, such as gravitational waves \cite{Graham2013,AGIS2008} or post-Newtonian effects \cite{GRAI}.  In parallel, a wide variety of formalisms for computing \ai phase shifts have been developed \cite{storey1994feynman,antoine2003exact,GRAI,tan2017calculating,Roura2020,ufrecht2020perturbative,Werner2024,Badurina2025,Roura2025,niehof2025finite} in both non-relativistic and relativistic settings.  However, to the best of our knowledge, all relativistic calculations so far have been perturbative and only carried out to finite order. 
Such approximate calculation methodologies can be computationally cumbersome \cite{Badurina2025}, can miss interesting physical effects of high perturbation order, and can pose difficulties of interpretation, since e.g. similar terms in a perturbation series may originate from multiple effects. 
Moreover, some formalisms introduce intermediate quantities which may not be independently measurable, which has resulted in previous work studying which aspects of \ai phase shift calculations are physical as opposed to artifacts of the computational method \cite{overstreet2021physically}.  Exact solutions for \ai phase shifts would have the potential to illuminate many of these issues, since an exact final phase shift expression exhibits no arbitrariness of decomposition, includes all possible effects, and may be analytically easier to work with than a perturbation series. 

In the present work, we use the covariant formalism \cite{GRAI} to calculate \ai phase shifts exactly in flat spacetime.  The case where the lasers driving atom optics pulses are in uniform hyperbolic motion can be considered a relativistic generalization of the non-relativistic setting of a uniform gravitational field, with the flatness of spacetime generalizing what would be classically described as a vanishing gravity gradient.  For specific choices of laser acceleration, this setting is equivalent to an \ai with lasers at fixed position in Rindler space \cite{niehof2025finite}.  We find that the kinematics of relativistic \ais are naturally adapted to a description in light-cone (LC) coordinates, which results in great simplification of the resulting recursive equations of motion, and much of the work of the present treatment consists in systematically expressing atom interferometer kinematics in LC coordinates.  All of our methods apply equally to either two-photon (Raman or Bragg) or single-photon (clock) \ais, though the analytical form of the resulting phase shifts can differ considerably.  

While our treatment of relativistic kinematics is exact, there are other approximations inherent to our approach, such as the semi-classical approximation and the short pulse approximation.  Both of these are standard in previous similar work \cite{GRAI}, and corrections to these approximations have been considered elsewhere \cite{antoine2006matter,jansen2007atom,glick2026feynman}.
Practical terrestrial \ai experiments are also significantly affected by spacetime curvature.  However, the flat space case captures the leading order and many sub-leading order effects, and it may provide an accurate foundation for perturbative treatments of curvature effects.  Moreover the mathematical structures which are evident in the flat space case are of interest in their own right, and provide useful intuition for the curved space case as well. 

Finite speed of light and laser chirping effects in atom interferometry have been considered before using perturbative frameworks \cite{le2007influence,cheng2015influence,tan2017calculating}.  Recent work has compared the influence of such effects in Raman and clock interferometers in Rindler space \cite{niehof2025finite}.  Our work extends these results by providing atom trajectories and interferometer phase shifts which are exact to all orders in relativistic parameters. 

The structure of this paper is as follows.  In \cref{sec_background} we recapitulate necessary background material on atom interferometry and relativistic kinematics.  In \cref{sec_1laser} we derive formulas for clock \ai phase shifts when all atom optics pulses are from a single direction.  In \cref{sec_2photon} we consider the simplest cases of Bragg and Raman (two-photon) \ais.  In \cref{sec_lmt} we treat clock \ais with atom optics pulses from two directions, and large momentum transfer (LMT) sequences in particular. 
In \cref{sec_off_shell} we treat off-shell transitions. 
In \cref{sec_discussion} we provide discussion of the results and directions for future work.  We use natural units $c=\hbar=1$ throughout. 

\section{Background} 
\label{sec_background}
This section summarizes background material on atom interferometry, relativistic kinematics, and atomic recoil kinematics which will be used often in the rest of this work.  

\subsection{Atom Interferometer Phase Shifts}
We follow the covariant formalism of \cite{GRAI} for computing \ai phase shifts.  The total \ai phase shift $\Delta\phi$ is decomposed as
\begin{equation}
\Delta\phi = \Delta\phi_{\text{prop}} + \Delta\phi_{\text{clock}} + \Delta\phi_{\text{sep}} + \Delta\phi_{\text{laser}}.
\end{equation}
The propagation phase $\Delta\phi_{\text{prop}}$ is the difference in proper time along the upper and lower arm trajectories, multiplied by the atomic ground state mass $m$, 
\begin{equation}
\Delta\phi_{\text{prop}} = m \int_{\text{upper}} \intd{\tau} - m\int_{\text{lower}}\intd{\tau},
\end{equation}
where $\intd{\tau}$ is differential proper time. 
The clock phase $\Delta\phi_{\text{clock}}$ is the difference in proper time spent in the excited state of the two arms, multiplied by the excited state energy $\omega_a$,
\begin{equation}
\Delta\phi_{\text{clock}} = \omega_a \int_{\substack{\text{upper}\\ \text{excited}}}\intd{\tau} - \omega_a \int_{\substack{\text{lower}\\ \text{excited}}}\intd{\tau}.
\end{equation}
Note that in \cite{GRAI}, the clock phase is included as part of the propagation phase.  We distinguish these two for convenience in organizing computations. 

The separation phase $\Delta\phi_{\text{sep}}$ in flat space is given by the inner product of the mean output port \mom $\bar{p}^\mu_f$ with the spatial separation $x^\nu_{\text{upper}} - x^\nu_{\text{lower}}$ of the two arms at the final beamsplitter,
\begin{equation}
\Delta\phi_{\text{sep}} = -\eta_{\mu\nu} \overline{p}_{f}^\mu \left(x^\nu_{\text{upper}} - x^\nu_{\text{lower}}\right). 
\end{equation}
where $\eta_{\mu\nu} = \text{diag}(1,-1)$ is the Minkowski metric. 

Finally the laser phase $\Delta\phi_{\text{laser}}$ is 
\begin{equation}
\Delta\phi_{\text{laser}} = \sum_{j, \text{ upper}} \pm \theta_j - \sum_{j, \text{ lower}} \pm \theta_j
\end{equation}
where the sum is over all laser-atom interactions, $\theta_j$ is the phase of the interacting light pulse, and the $\pm$ sign is $+$ if the interaction results in the atom absorbing a photon and $-$ if it results in the atom emitting a photon. 

\subsection{Kinematics in Light Cone Coordinates}
\label{sec_lc}
Flat 1+1 dimensional spacetime with Minkowski coordinates $\invec{t}{x}$ and metric $\eta_{\mu\nu} = \inmat{1}{0}{0}{-1}$ can be parametrized by light-cone (LC) coordinates $\invec{r^+}{r^-}=\invec{t+x}{t-x}$, with the metric taking the form $\frac{1}{2}\inmat{0}{1}{1}{0}$.  We refer to $r^+$ and $r^-$ respectively as the \lcp and \lcm coordinates of the corresponding spacetime point.  A future-directed, time-like unit vector has components $\invec{\cosh(\beta)}{\sinh(\beta)}$ in Minkowski coordinates or $\invec{e^{\beta}}{e^{-\beta}}$ in LC coordinates, where $\beta$ is the boost angle (rapidity) with respect to a particle at rest in the frame.  Coordinate velocity $v$ is related to $\beta$ by $v=\tanh(\beta)$, $\beta=\frac{1}{2}\log\big(\frac{1+v}{1-v}\big)$.

The great advantage of LC coordinates emerges when we consider atom-photon interactions.  An upwards-moving atom optics pulse travels a line of fixed LC coordinate $r^-$, and a downwards-moving pulse travels a line of fixed $r^+$.  For an atom with initial LC position $\invec{r^+_0}{r^-_0}$ and velocity $\invec{e^{\beta}}{e^{-\beta}}$, the proper time which the atom must travel to interact with a light pulse of fixed LC coordinate $\ell^\pm$ is 
\begin{equation}
\label{eq_s}
s = \frac{\ell^\pm - r^\pm_0}{e^{\pm \beta}},
\end{equation}
and the other LC coordinate of the atom at this interaction point will be
\begin{equation}
\label{eq_r}
r^\mp_1 = r^\mp_0 + s e^{\mp\beta}.
\end{equation}

A laser in uniform hyperbolic motion (i.e. undergoing uniform proper acceleration) with acceleration $\alpha$ has trajectory
\begin{align}
\dvec{t(\tau)}{x(\tau)} & = \dvec{t_0}{x_0} + \frac{1}{\alpha}\dvec{\sinh(\alpha \tau + \kappa) - \sinh(\kappa)}{\cosh(\alpha\tau + \kappa) - \cosh(\kappa)} \label{eq_hypertx} \\
\dvec{r^+(\tau)}{r^-(\tau)} & = \dvec{r^+_i}{r^-_i} + \frac{1}{\alpha}\dvec{e^{\alpha\tau + \kappa}}{-e^{-\alpha\tau-\kappa}} \label{eq_hyperlc}
\end{align}
in Minkowski or LC coordinates, respectively.  Here $\tau$ is the laser's proper time, $\kappa$ is the laser's rapidity at proper time $\tau=0$, $t_0$ and $x_0$ are the laser's Minkowski coordinates at $\tau=0$, and $r^\pm_i$ are related to $t_0, x_0$ by $r^\pm_i=t_0\pm x_0\mp e^{\pm\kappa}/\alpha$. 

\subsection{Single-Photon Recoil Boosts}
\label{sec_boosts}
All vectors in this subsection will be presented in Minkowski coordinates.  Photon \moma have the form $\invec{\omega}{\pm\omega}$, with sign $+$ for upwards moving and $-$ for downwards moving trajectories.  When an atom of mass $m$ initially at rest in the lab frame absorbs a resonant photon of \mom $\invec{\omega}{\pm\omega}$, the atom's final \mom is $\invec{m+\omega}{\pm\omega}$, and the atom's rapidity (i.e. boost angle) is thus increased by \cite{swan2025atom}
\begin{align}
\label{eq_rest_recoil_boost}
\beta_r & = \tanh^{-1}\left(\frac{\pm\omega}{m+\omega}\right) = \pm \ln\sqrt{1+\frac{2\omega}{m}}.
\end{align}
We refer to $\beta_r$ as the recoil boost.  If the atom is moving with respect to the reference frame with initial rapidity $\beta_0$, then the boost from a photon of frequency $\omega$ can be computed by first transforming the photon \mom to the atom's rest frame and then applying \cref{eq_rest_recoil_boost}, with the result
\begin{equation}
\beta_r = \tanh^{-1}\left(\frac{\pm e^{\mp\beta_0}\omega}{m+e^{\mp\beta_0}\omega}\right).
\end{equation}
Similarly, if the incoming photon causes stimulated emission from an atom initially in an excited state of \mom $\invec{m+\omega_a}{0}$, where $\omega_a$ is the excited state frequency, the atom experiences a boost
\begin{equation}
\beta_r = \tanh^{-1}\left(\frac{\mp e^{\mp\beta_0}\omega}{m+\omega_a-e^{\mp\beta_0}\omega}\right).
\end{equation}
Note that due to recoil shift the frequency of a resonant photon $\omega_{\text{res}}$ is not equal to the frequency $\omega_a$ of the atomic excited state it drives.  The relationship between these is determined from the condition that an atom initially at rest with \mom $\invec{m}{0}$, after absorbing a photon of \mom $\invec{\omega_{\text{res}}}{\omega_{\text{res}}}$, has \mom with norm $m+\omega_a$.  This fixes 
\begin{equation}
\label{eq_recoil_shift}
\omega_{\text{res}} = \omega_a \left(1 + \frac{\omega_a}{2m}\right).
\end{equation}
We speak of atom optics pulses of this frequency as being on-shell.  Note that from \cref{eq_recoil_shift} the quantity $e^{\beta_r}$ (the \lcp component of the atom's velocity, which will often appear later) for an on-shell photon traveling upwards is given by
\begin{equation}
\label{eq_ebr}
e^{\beta_r} = 1+\frac{\omega_a}{m}.
\end{equation}

When we consider two-photon \ais, we will require formulas analogous to \cref{eq_rest_recoil_boost,eq_recoil_shift,eq_ebr} for this case.  Since there are several sub-cases and the algebra is more involved, we provide these formulas in \cref{app_boosts}.

\section{Single-Direction Clock Atom Interferometers}
\label{sec_1laser}

\begin{figure}[t]
\centering
    \includegraphics[width=0.99\linewidth]{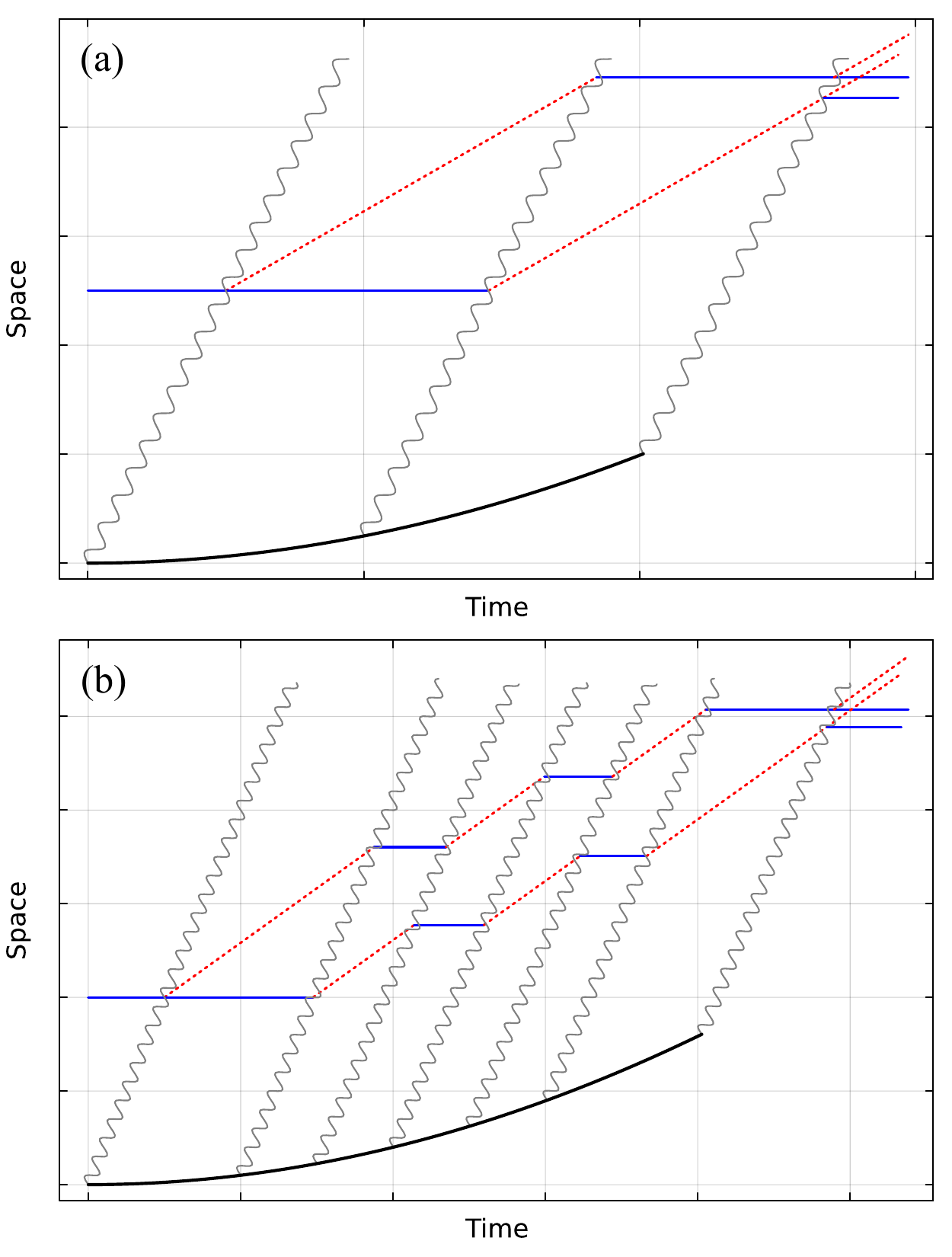}\vspace{-0.38cm}
    \caption{Relativistic clock atom interferometers with atom optics pulses from a single direction.  Solid blue lines and dotted red lines indicate atom trajectories with the atom in the ground state or excited state, respectively.  The solid black line indicates the laser trajectory, which here is depicted as hyperbolic.  Wavy grey lines are atom optics photon trajectories.  (a) Standard Mach-Zehnder sequence.  (b) Generic pulse sequence.}
\label{fig_mz} 
\end{figure}

In this section we consider the case of a clock atom interferometer where all atom optics pulses travel in a single direction.  We further suppose that all pulses are exactly on-shell.  A typical Mach-Zehnder \ai of this type is shown in \cref{fig_mz}(a), and a more general such \ai is shown in \cref{fig_mz}(b).  Each $\pi$ pulse toggles the state of both arms of the interferometer.  Whenever an arm is in the ground state, its rapidity is $\beta_0$, and whenever it is in the excited state its rapidity is $\beta_0 + \beta_r$.

The atom optics pulses have \lcm coordinates $\ell_j^-$ where $j=1,\dots, N$, with $j=1$ and $j=N$ being $\pi/2$ pulses and all others being $\pi$ pulses.  We suppose the atom has initial LC coordinates $\invec{r^+_0}{r^-_0}$ and boost angle $\beta_0$.  The positions of the upper and lower arms of the interferometer at their interactions with pulse $j$ are denoted $\invec{r^+_{j,p}}{r^-_{j,p}}$, where $p=1$ for the upper arm and $p=0$ for the lower arm.  The velocity of arm $p$ between pulse $j$ and $j+1$ is $\invec{v^+_{j,p}}{v^-_{j,p}} = \invec{e^{\beta_0}}{e^{-\beta_0}}$ when $p+j$ is odd and $\invec{v^+_{j,p}}{v^-_{j,p}} = \invec{e^{\beta_0+\beta_r}}{e^{-\beta_0-\beta_r}}$ when $p+j$ is even.  The proper time elapsed by arm $p$ in traveling from pulse $j$ to pulse $j+1$ is
\begin{equation}
\label{eq_sj1}
s_{j,p} = \begin{cases} 
\frac{\ell^-_{j+1} - \ell^-_j}{e^{-\beta_0}} & j+p\text{ odd} \\
\frac{\ell^-_{j+1} - \ell^-_j}{e^{-\beta_0-\beta_r}} & j+p\text{ even} \\
\end{cases}
\end{equation}
by \cref{eq_s}.  We may immediately compute the propagation phase 
\begin{multline}
\label{eq_prop1}
\Delta\phi_{\text{prop}} = m\left(\sum_{j=1}^{N-1} s_{j,1} - \sum_{j=1}^{N-1} s_{j,0}\right) \\ = m e^{\beta_0} \left(e^{\beta_r}-1\right)\sum_{j=1}^{N-1} (-1)^{j+1}\left(\ell^-_{j+1} - \ell^-_j\right)
\end{multline}
and the clock phase 
\begin{multline}
\label{eq_clock1}
\Delta\phi_{\text{clock}} = \omega_a \left( \sum_{\substack{j=1 \\ j\text{ odd}}}^{N-1} s_{j,1} - \sum_{\substack{j=2 \\ j\text{ even}}}^{N-1} s_{j,0} \right) \\
= \omega_a e^{\beta_0 + \beta_r} \sum_{j=1}^{N-1} (-1)^{j+1} \left(\ell^-_{j+1} - \ell^-_j\right).
\end{multline}
To compute the separation phase, we first compute the separation $\Delta r^+ := r^+_{N,1} - r^+_{N,0}$ at the final beamsplitter.  From \cref{eq_r,eq_sj1}, we find 
\begin{multline}
r^+_{j+1,1} - r^+_{j+1,0} = r^+_{j,1} - r^+_{j,0} \\ 
+ e^{2\beta_0} \left(e^{2\beta_r}-1\right) (-1)^{j+1} \left(\ell^-_{j+1} - \ell^-_j\right).
\end{multline}
Together with the initial condition $r^+_{1,1} - r^+_{1,0} = 0$, this gives 
\begin{equation}
\Delta r^+ = e^{2\beta_0}\left(e^{2\beta_r}-1\right) \sum_{j=1}^{N-1} (-1)^{j+1} \left(\ell^-_{j+1} - \ell^-_j\right)
\end{equation}
Then the \ai separation phase is given by the inner product of the final separation vector $\invec{\Delta r^+}{0}$ with the lower output port \mom vector, which has LC coordinates $m\invec{e^{\beta_0}}{e^{-\beta_0}}$, yielding
\begin{multline}
\label{eq_clock_sep}
\Delta\phi_{\text{sep}} = - \frac{1}{2}\Delta r^+ m e^{-\beta_0} \\ 
= -\frac{m}{2} e^{\beta_0} \left(e^{2\beta_r}-1\right) \sum_{j=1}^{N-1} (-1)^{j+1} \left(\ell^-_{j+1} - \ell^-_j\right),
\end{multline}
with the factor of $\frac{1}{2}$ coming from the LC metric.  Equivalently, we could have used the upper output port \mom vector $(m+\omega_a)\invec{e^{\beta_0+\beta_r}}{e^{-\beta_0-\beta_r}}$; since this differs from the lower port \mom by an upwards null vector and the atom separation is also an upwards null vector, the result will be the same.

Together the propagation, clock, and separation phases account for all contributions to the total \ai phase which depend on atom kinematics.  The remaining contribution is from the laser phase, which requires additional problem data to describe the laser phase evolution.  We will discuss laser phase below.  Excluding this for now, the remaining phase can be written 
\begin{multline}
\label{eq_kin1}
\Delta\phi_{\text{prop}} + \Delta\phi_{\text{clock}} + \Delta\phi_{\text{sep}} = \\
e^{\beta_0}\left(\omega_a e^{\beta_r} - \frac{m}{2} \left(e^{\beta_r}-1\right)^2\right) \sum_{j=1}^{N-1} (-1)^{j+1} \left(\ell^-_{j+1} - \ell^-_j\right) \\
= e^{\beta_0} \omega_a \left(1+\frac{\omega_a}{2m}\right)\sum_{j=1}^{N-1} (-1)^{j+1} \left(\ell^-_{j+1} - \ell^-_j\right),
\end{multline}
where we have used \cref{eq_ebr} in the final equality. 
We see that the entire effect of the pulse sequence is described by the sum
\begin{equation}
A := \sum_{j=1}^{N-1} (-1)^{j+1} \left(\ell^-_{j+1} - \ell^-_j\right),
\end{equation}
with the rest of the kinematic phase being given by the atom's intrinsic properties and initial conditions.  For conciseness, we refer to $A$ as the ``record'' associated with a given sequence. 

\begin{figure}[t]
\centering
    \includegraphics[width=0.99\linewidth]{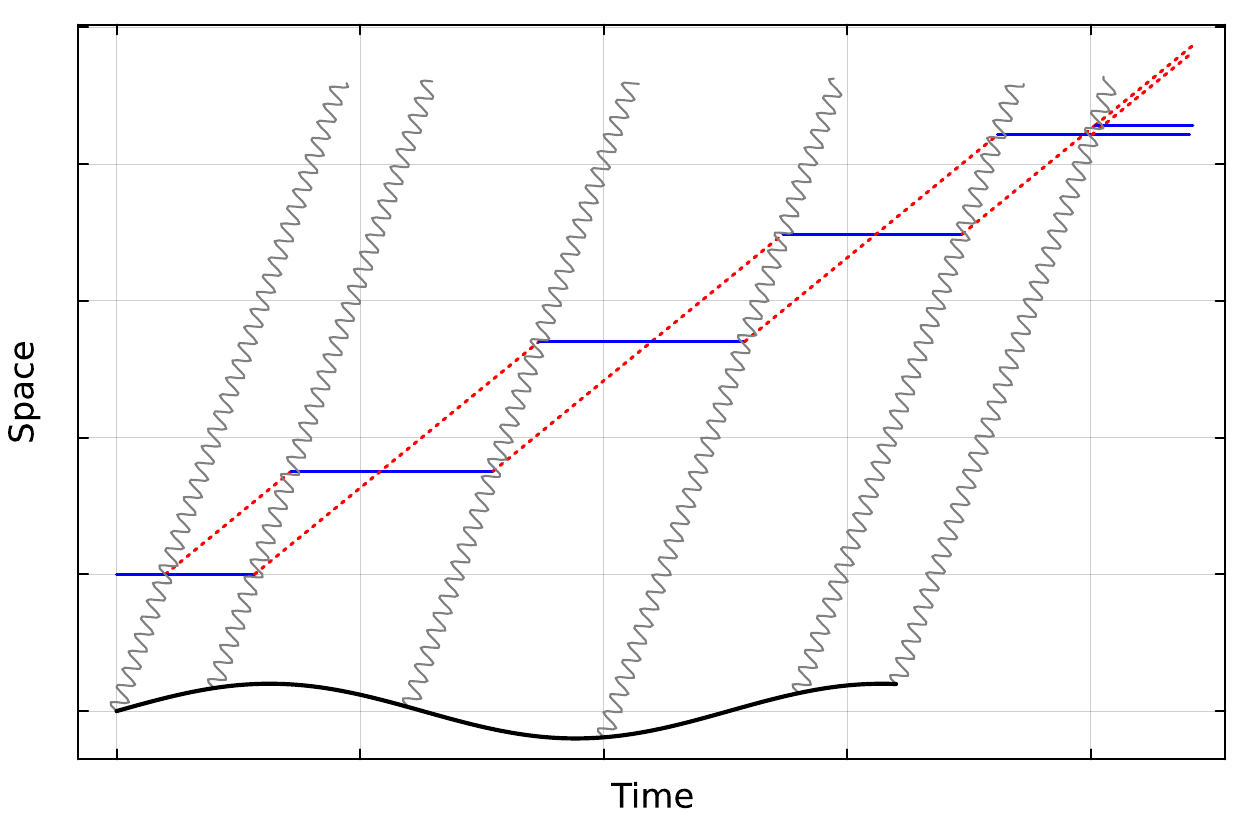}\vspace{-0.38cm}
    \caption{$Q=4$ resonant atom interferometer.  The upper and lower arms of this interferometer cross between each successive $\pi$ pulse, which results in enhanced sensitivity to oscillating signals commensurate with the pulse frequency.}
\label{fig_res1} 
\end{figure}

We now compute records for some explicit laser trajectories of interest. Consider first a Mach-Zehnder sequence for a laser in hyperbolic motion as described by \cref{eq_hyperlc} with acceleration $\alpha = g$ and initial boost $\kappa=0$.  The laser fires pulses at proper times $0$, $T$, and $2T$.  The fixed LC coordinates of these pulses are $\ell^-_1=r^-_i - 1/g$, $\ell^-_2=r^-_i - e^{-gT}/g$, and $\ell^-_3=r^-_i - e^{-2gT}/g$, yielding 
\begin{equation}
\label{eq_AMZ}
A_{\text{MZ}} = \frac{\left(e^{-gT}-1\right)^2}{g}.
\end{equation}
Another interesting pulse sequence is a resonant one \cite{graham2016resonant}, such as that shown in \cref{fig_res1}.  For analytical simplicity, we take the laser trajectory to be given as a function of coordinate time by
\begin{equation}
\label{eq_sine_laser}
x(t) = x_0 + a\sin(\Omega t)
\end{equation}
\begin{figure}[t]
\centering
    \includegraphics[width=\figwidth]{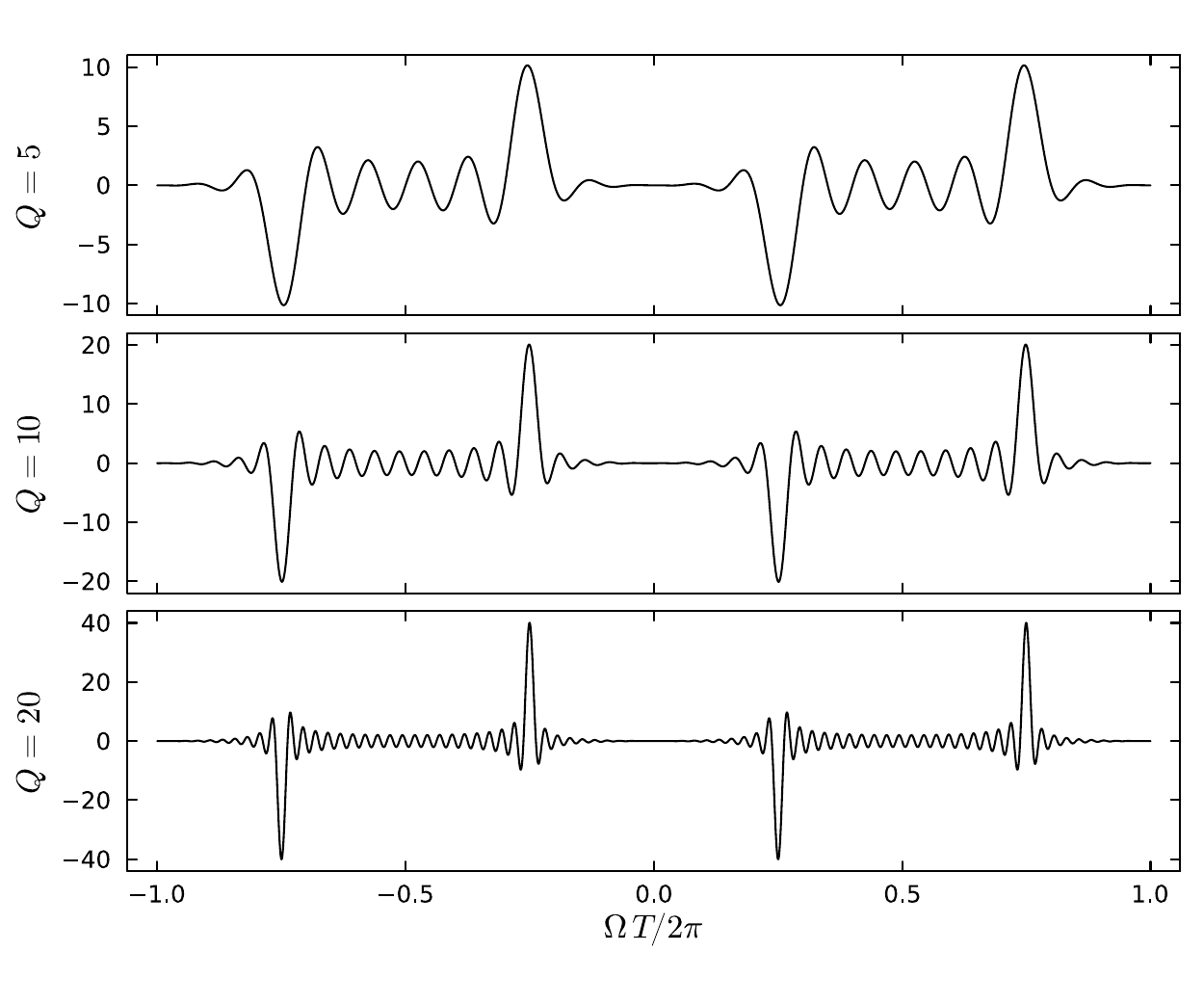}\vspace{-0.38cm}
    \caption{Resonant pulse sequence record $A_{\text{res}}$ vs. $\Omega T$ for three values of $Q$, with amplitude $a=1$.  Note the different vertical axes for each plot.  The extremal values of the record are $\pm 2Q$, which is the resonant enhancement effect.}
\label{fig_record} 
\end{figure}
We suppose the opening beamsplitter pulse leaves the laser at coordinate time $t=0$, followed by $Q$ successive $\pi$ pulses at coordinate times $t=T,3T,\dots,(2Q-1)T$ and a final beamsplitter at coordinate time $2QT$.  The resulting record is then
\begin{equation}
A_{\text{res}} = (-1)^Q a \sin(2\Omega Q T)\left(\sec(\Omega T)-1\right).
\end{equation}
This record is plotted for several values of $Q$ in \cref{fig_record}.  The maximum value of $A_{\text{res}}$ is $2Qa$, which occurs at $\Omega T = \pi\left(\frac{1}{2}+M\right)$ for integer $M$.

\subsection{Laser Phase}
The laser phase is conceptually distinct from the other \ai phase terms, as it requires a specification of intrinsic properties of the laser which are otherwise irrelevant to the \ai, such as e.g. laser phase noise or deliberately applied phase shifts.  Denoting the phase of the laser at pulse $j$ by $\theta_j$, the associated contribution to the \ai phase is in general
\begin{align}
\Delta\phi_{\text{laser}} & = \theta_1 + (-1)^{N+1} \theta_N + \sum_{j=2}^{N-1} 2(-1)^{j+1} \theta_j \\
& = - \sum_{j=1}^{N-1} (-1)^{j+1}\left(\theta_{j+1} - \theta_{j}\right) .
\label{eq_laser_phase}
\end{align}

The mathematically simplest model for the laser phase is to assume that each pulse has zero phase, which is physically reasonable, though not what is typically realized in an experiment. We can imagine this case as arising from an ideal laser system which has a knob to set the frequency and a button to fire a pulse, with each pulse being a perfect truncated sine wave starting from zero phase.  This model captures all relevant effects for a gradiometric configuration \cite{Graham2013} and also serves as a base case to which arbitrary laser phases can be added.  With this laser phase prescription, the kinematic phase of \cref{eq_kin1} is the entire \ai phase.  

For a Mach-Zehnder \ai with zero laser phase, from \cref{eq_kin1,eq_AMZ}, the full phase shift is then 
\begin{align}
\label{eq_MZ_L0}
\Delta\phi & = e^{\beta_0} \omega_a \left(1+\frac{\omega_a}{2m}\right) \frac{\left(e^{-gT}-1\right)^2}{g} \\
& = \omega_a \left(1+\frac{\omega_a}{2m}\right) \sqrt{\frac{1+v_0}{1-v_0}} \frac{\left(e^{-gT}-1\right)^2}{g} \label{eq_MZ_L0_v} \\
& = \omega_a g T^2 - \omega_a g^2 T^3 + \omega_a v_0 g T^2 + \frac{\omega_a^2}{2m}g T^2 + \cdots
\end{align}
To facilitate comparison with other literature \cite{GRAI,Werner2024,niehof2025finite}, in the second line we have expressed $e^{\beta_0}$ in terms of the atom's initial coordinate velocity in the lab frame $v_0$, and in the third line we have given the lowest order terms of a series expansion about $T=0$ and $v_0=0$. 

Returning now to generic pulse sequences, a more common model for the laser phase in practice is that it arises from a laser which oscillates continually at either a fixed frequency $\omega_{\ell}$ or at a chirped frequency $\omega_{\ell}(\tau)$, where $\tau$ is the laser's proper time.  In the case where the laser is in uniform hyperbolic motion as in \cref{eq_hyperlc} with $\alpha=g$ and $\kappa =0$, an upwards traveling photon of frequency $\omega_{\ell}(\tau)$ in the laser's rest frame has Doppler-shifted frequency $e^{g\tau} \omega_{\ell}(\tau)$ in the lab frame.  Thus if such a laser is to source atom optics pulses which are precisely on-resonance, the laser frequency must satisfy
\begin{equation}
\omega_\ell(\tau) = \omega_\text{res} e^{\beta_0} e^{-g\tau}.
\end{equation}
If the laser phase evolution results from directly integrating this frequency schedule, then the resulting laser phase contribution to the \ai phase shift is given by 
\begin{multline}
\label{eq_laser_phase_1}
\Delta\phi_{\text{laser}} = \frac{1}{g}\omega_a \left(1+\frac{\omega_a}{2m}\right) e^{\beta_0} \\
\times \sum_{j=1}^{N-1} (-1)^{j+1} \left(e^{-g\tau_{j+1}} - e^{-g\tau_{j}}\right).
\end{multline}
Comparing to \cref{eq_kin1,eq_hyperlc} for the same laser trajectory, we see that in this case
\begin{align}
\Delta\phi_{\text{laser}} & = - e^{\beta_0} \omega_a \left(1 + \frac{\omega_a}{2m}\right) A \\ 
& = -\left(\Delta\phi_{\text{prop}} + \Delta\phi_{\text{clock}} + \Delta\phi_{\text{sep}}\right) .
\end{align}
We thus see that the total phase shift $\Delta\phi$ exactly vanishes.  The fact that the laser phase should at least partially cancel the kinematic phase terms in this case is unsurprising: By the equivalence principle, a Mach-Zehnder \ai has zero phase when the laser is freely falling with the atom, and chirping the laser to keep it resonant with the atom simulates this scenario up to the small pulse timing deviations caused by time dilation of the laser.  The fact that the total phase exactly vanishes in this case may seem surprising at first, since e.g. the pulse timing is not uniform in the lab frame and the interferometer does not fully close at the final beamsplitter (unlike in a non-relativistic treatment).  However, the locking of atom and laser phases imposed by the on-shell condition ensures that the total interferometer phase is zero even in the presence of arbitrary pulse timing deviations. 

\section{Two-Photon Atom Interferometers}
\label{sec_2photon}

\begin{figure}[t]
\centering
    \includegraphics[width=0.99\linewidth]{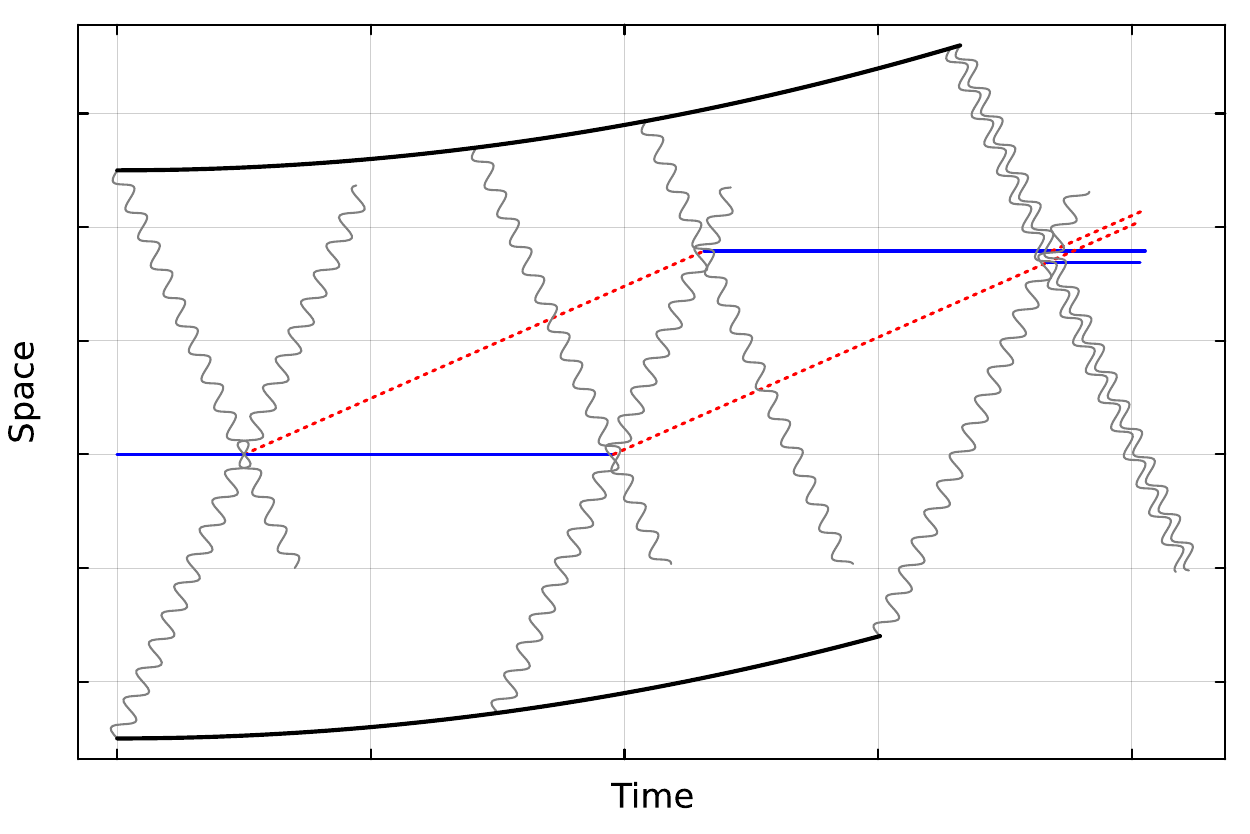}\vspace{-0.38cm}
    \caption{Mach-Zehnder Raman atom interferometer.  Solid blue and dotted red lines now correspond to the two atomic hyperfine states of the Raman transition. The upper (passive) laser is assumed to be always on, while the lower (active) laser determines the atom-light interaction points which then pick out respective photons from the upper laser.}
\label{fig_mz_raman} 
\end{figure}

We consider next the simplest case of two-photon \ais using Raman or Bragg atom optics, which closely parallels the single-direction clock \ai calculation.  The two-photon boost angle formulas we will need are presented in \cref{app_boosts}. We now have a lower laser following a hyperbolic trajectory with initial spatial position $-L/2$, initial rapidity zero, and acceleration $\alpha_-$, and an upper laser following a hyperbolic trajectory with initial spatial position $L/2$, initial rapidity zero, and acceleration $\alpha_+$. We suppose that the upper laser is passive and always on, while the lower laser is the control laser and thus is pulsed and determines the coordinates of all atom-light interaction points \cite{GRAI}. 
A special case of interest is that where the lasers are rigidly connected, in the sense of Born rigidity \cite{born1909theorie,born1909dynamik}.  This models a situation such as two lasers fixed to opposite ends of the same rocket, or equivalently lasers at fixed positions in Rindler coordinates, and constrains their proper accelerations to satisfy 
\begin{equation}
\frac{1}{\alpha_+} - \frac{1}{\alpha_-} = L.
\end{equation}
We will not generally specialize to this case, however, as it results in no significant simplification of any expressions. 

We suppose that each laser emits a chirped frequency according to 
\begin{equation}
\omega_\pm(\tau) = \omega_{\pm,0} e^{\gamma_\pm \tau}
\end{equation}
in the respective laser's instantaneous rest frame, where $\gamma_\pm$ is a chirp rate.
We suppose that the initial frequencies $\omega_{\pm,0}$ satisfy the on-shell condition \cref{eq_on_shell1}.  However, at later times this condition will in general be violated except in the special case $\gamma_\pm = \pm \alpha_\pm$ so that the chirp exactly cancels the atom optic Doppler shift due to the laser motion.  In this section we therefore specialize to this case, deferring the case of general chirp to \cref{sec_off_shell_raman}.  The atom dynamics then has exactly the same form as for a single-direction clock interferometer, with the boost angle $\beta_r$ now given by \cref{eq_os_raman_boost}.  The kinematic contributions to the phase shift therefore have the same form as \cref{eq_kin1}, with $\omega_a$ replaced by the hyperfine splitting $\Delta\omega_a$ (which is zero for a Bragg \ai).  The phase contribution from the lower laser also exactly parallels the clock \ai case.  If the laser phase arises purely from the frequency chirp (as in \cref{eq_laser_phase_1}), the phase $\phi_{\pm}$ of the upper ($+$) or lower ($-$) laser satisfies
\begin{equation}
\label{eq_resonant_chirped_phase}
\phi_{\pm}(\tau) = \pm \frac{\omega_{\pm,0}}{\alpha_\pm} e^{\pm\alpha_\pm\tau} 
\end{equation}
up to an overall constant which will drop out.  The phase of each laser is thus equal to its \lcpm coordinate times $\omega_{\pm,0}$ (again modulo an irrelevant constant).  Analogously to \cref{eq_laser_phase_1}, the lower laser phase contribution is thus
\begin{equation}
\Delta\phi_{\text{laser},-} = -\omega_{-,0} A.
\end{equation}

Since the upper laser's \lcp coordinate at the time of emission for a pulse which will interact with the atom coincides with the atom's \lcp coordinate at the interaction point, the contribution of the upper laser's phase to the overall \ai phase is given by a sum of terms of the form $\pm \omega_{+,0} \, r^+_{j,p}$, with a $+$ sign when the lower arm gains momentum or the upper arm loses momentum, and a $-$ in the remaining cases.  In the case of a Mach-Zehnder \ai, this becomes, for the lower output port,
\begin{align}
\Delta\phi_{\text{laser},\,+} & = \omega_{+,0} \left(r^+_{2,1} - r^+_{1,1} + r^+_{2,0} - r^+_{3,0}\right) \\
& = \omega_{+,0} e^{2\beta_0 + 2\beta_r} \left( - \ell^-_{1} + 2 \ell^-_{2} - \ell^-_{3} \right) \\
& = \omega_{+,0} e^{2\beta_0 + 2\beta_r} A_{\text{MZ}},
\end{align}
where we have used \cref{eq_r,eq_sj1,eq_AMZ}. One can show more generally that an analogous relationship holds for arbitrary pulse sequences and records $A$ provided (as in the single-direction clock case) the atom only toggles between two momentum states.  The full phase shift for such a Bragg or Raman \ai is then 
\begin{multline}
\Delta\phi = \\ e^{\beta_0} A \Big[
e^{\beta_r}\Delta\omega_a - \frac{m}{2}\left(e^{\beta_r} \! - \! 1\right)^2  - \omega_{-,0} e^{-\beta_0} + \omega_{+,0} e^{\beta_0 + 2\beta_r}\Big].
\end{multline}
Using formulas of \cref{app_boosts}, one can show that the bracketed terms in this expression completely cancel.  Thus as in the case of a resonantly chirped clock \ai, we find that the phase shift of a resonantly chirped Raman or Bragg \ai is precisely zero.  We will find more non-trivial expressions for Raman or Bragg phase shifts when we consider off-shell transitions in \cref{sec_off_shell_raman} below.

\section{Large Momentum Transfer}
\label{sec_lmt}

\begin{figure}[t]
\centering
    \includegraphics[width=0.99\linewidth]{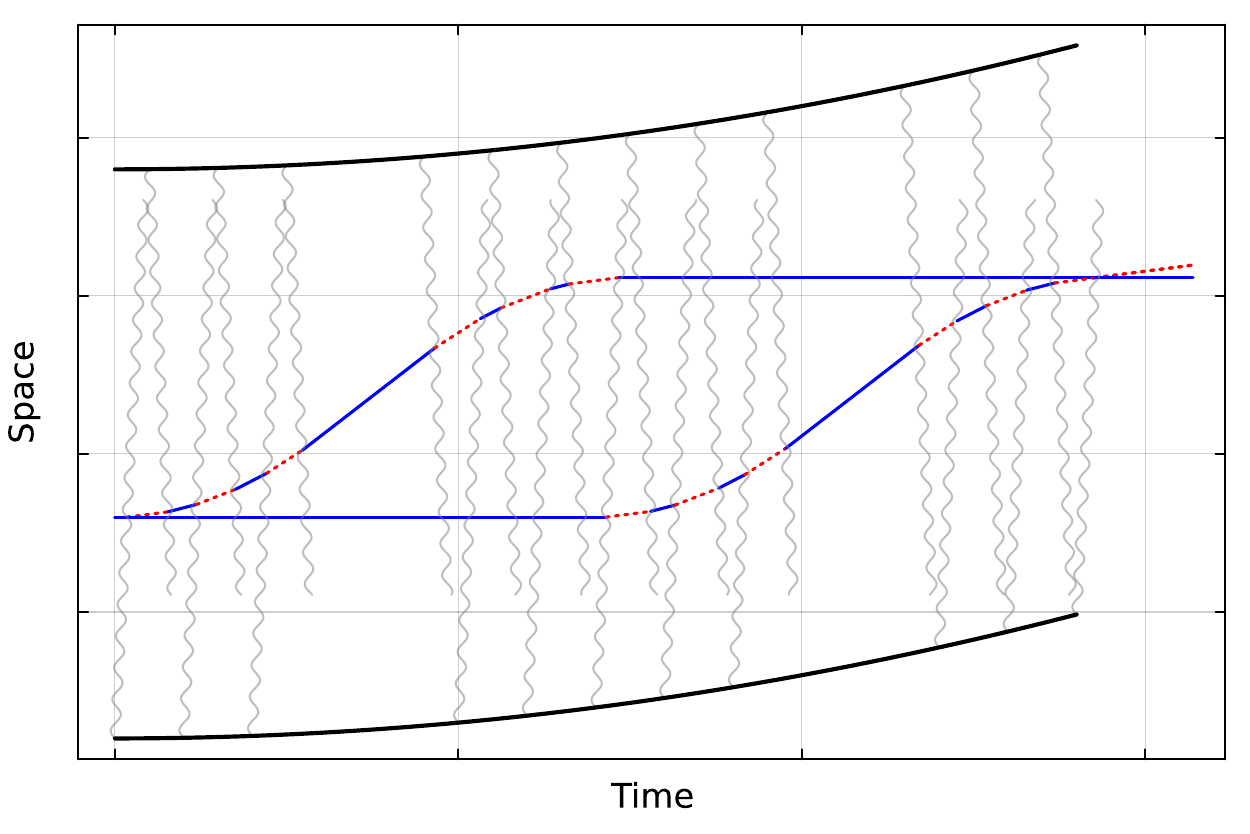}\vspace{-0.38cm}
    \caption{$N=3$ LMT narrowband Mach-Zehnder clock atom interferometer.  The single-photon atom optics of a standard Mach-Zehnder are replaced by $2N$ atom optics.}
\label{fig_lmt} 
\end{figure}

We next investigate a large momentum transfer (LMT) Mach-Zehnder clock atom interferometer, as depicted in \cref{fig_lmt}.  The maximum momentum separation of the wavepackets in this interferometer is $2N$ photon recoils; the case of an odd number of photon recoils can be treated similarly.  Full algebraic details for this derivation are available in \cite{math_supp}. 

We suppose that the interferometer begins with a $\pi/2$ pulse from the lower laser, followed by $(2N-1)$ $\pi$ pulses from alternating directions, starting with the upper laser.  This sequence of $2N$ pulses forms an effective LMT beamsplitter, analogous to a standard $\pi/2$ pulse, but with $2N$ times larger recoil boost.  After a drift time, another sequence of $4N-1$ alternating direction $\pi$ pulses, starting with the upper laser, serve to bring the upper arm back to its initial \mom while accelerating the lower arm.  After another drift period, a sequence of $2N-1$ $\pi$ pulses, starting from the upper laser, followed by a $\pi/2$ pulse from the lower laser, act as another LMT beamsplitter to recombine the two arms.  We assume that all pulses are frequency selective such that only the $\pi/2$ pulses and the central $\pi$ pulse interact with both arms; all other $\pi$ pulses in the first half of the sequence interact only with the upper arm, while all those in the latter half of the sequence interact only with the lower arm.  This corresponds to the ``narrowband'' \ai described in \cite{jiang2025cumulative}. 

We begin by analyzing in a general way the effect on one arm of the interferometer of a sequence of pulses from alternating direction, as in the opening LMT beamsplitter or one half of the central LMT mirror sequence.  Suppose that the first pulse comes from below and that the atom is initially in its ground state with LC coordinates $\invec{r_0^+}{r_0^-}$ and with initial velocity $\invec{v^+_0}{v^-_0} = \invec{e^{\beta_0}}{e^{-\beta_0}}$.  We label the laser pulses by $j=1,\dots,2N$, with odd pulses coming from the lower laser and even pulses from the upper laser.  The fixed coordinates of these pulses are denoted $\ell^\pm_j$, with $+$ superscript for even $j$ and $-$ superscript for odd $j$.  We denote the LC coordinates of the atom at the point where it interacts with pulse $j$ by $\invec{r^+_j}{r^-_j}$.  We suppose all laser pulses are on-shell, so that each pulse augments the atom's rapidity by $\beta_r$ (see \cref{eq_ebr}).  Thus the atom's velocity immediately after pulse $j$ is $\invec{v^+_j}{v^-_j} = \invec{e^{\beta_0+j\beta_r}}{e^{-\beta_0-j\beta_r}}$.

From \cref{eq_s}, the proper time elapsed by the atom in traveling from $\invec{r^+_j}{r^-_j}$ to $\invec{r^+_{j+1}}{r^-_{j+1}}$ is given by 
\begin{equation}
\label{eq_s0_lmt}
s_j = \begin{cases}
\frac{\ell^+_{j+1} - r^+_j}{v^+_j} & j\text{ odd} \\
\frac{\ell^-_{j+1} - r^-_j}{v^-_j} & j\text{ even}
\end{cases}
\end{equation}
and the atom's position is then given by
\begin{align}
\label{eq_rp_lmt}
r^+_{j+1} & = \begin{cases}
\ell^+_{j+1} & j\text{ odd} \\
\ell^+_{j} + s_j v^+_j & j\text{ even}
\end{cases} \\
\label{eq_rm_lmt}
r^-_{j+1} & = \begin{cases}
\ell^-_{j+1} & j\text{ even} \\
\ell^-_{j} + s_j v^-_j & j\text{ odd}
\end{cases}
\end{align}
for $j \geq 1$.  For convenience, we define $\ell^+_0 := r^+_0$, so that \cref{eq_rp_lmt} holds for $j=0$ as well.  
Inserting \cref{eq_rp_lmt,eq_rm_lmt} into \cref{eq_s0_lmt} gives a recursion for $s_j$ alone,
\begin{equation}
\label{eq_s1_lmt}
s_j = \begin{cases}
\frac{\ell^+_{j+1} - \left(\ell^+_{j-1} + s_{j-1} v^+_{j-1}\right)}{v^+_j} & j\text{ odd} \\
\frac{\ell^-_{j+1} - \left(\ell^-_{j-1} + s_{j-1} v^-_{j-1}\right)}{v^-_j} & j\text{ even},
\end{cases}
\end{equation}
valid for $j\geq 1$.
Iterating this recursion once yields
\begin{equation}
\label{eq_s2_lmt}
s_j = \begin{cases}
\frac{\ell^+_{j+1} - \ell^+_{j-1}}{v^+_j} - \frac{v^+_{j-1}}{v^+_j} \frac{\ell^-_{j} - \ell^-_{j-2}}{v^-_{j-1}} + \frac{v^+_{j-1} v^-_{j-2}}{v^+_j v^-_{j-1}} s_{j-2} & j\text{ odd} \\
\frac{\ell^-_{j+1} - \ell^-_{j-1}}{v^-_j} - \frac{v^-_{j-1}}{v^-_j} \frac{\ell^+_{j} - \ell^+_{j-2}}{v^+_{j-1}} + \frac{v^-_{j-1} v^+_{j-2}}{v^-_j v^+_{j-1}} s_{j-2} & j\text{ even}
\end{cases}
\end{equation}
for $j \geq 2$.
Our hypothesis that all pulses are on resonance implies that 
$$
1 = \frac{v^+_{j-1} v^-_{j-2}}{v^+_j v^-_{j-1}} = \frac{v^-_{j-1} v^+_{j-2}}{v^-_j v^+_{j-1}}, 
$$
and thus, employing our expression for $v^\pm_j$, the recursion for $s_j$ becomes
\begin{equation}
s_j = \begin{cases}
\frac{\ell^+_{j+1} - \ell^+_{j-1}}{e^{\beta_0+j\beta_r}} - \frac{\ell^-_{j} - \ell^-_{j-2}}{e^{-\beta_0-(j-2)\beta_r} } + s_{j-2} & j\text{ odd} \\
\frac{\ell^-_{j+1} - \ell^-_{j-1}}{e^{-\beta_0-j\beta_r}} -  \frac{\ell^+_{j} - \ell^+_{j-2}}{e^{\beta_0 + (j-2)\beta_r}} + s_{j-2} & j\text{ even}.
\end{cases}
\end{equation}
This is easily solved to give 
\begin{equation}
\label{eq_lmt_s_sol}
s_j = \begin{cases}
\sum\limits_{\substack{k=3 \\ k\text{ odd}}}^j \frac{\ell^+_{k+1} - \ell^+_{k-1}}{e^{\beta_0+k\beta_r}} - \sum\limits_{\substack{k=2 \\ k\text{ even}}}^{j-1}\frac{\ell^-_{k+1} - \ell^-_{k-1}}{e^{-\beta_0-(k-1)\beta_r} } + s_{1} & j\text{ odd} \\
\sum\limits_{\substack{k=2 \\ k\text{ even}}}^j \frac{\ell^-_{k+1} - \ell^-_{k-1}}{e^{-\beta_0-k\beta_r}} - \sum\limits_{\substack{k=1 \\ k\text{ odd}}}^{j-1} \frac{\ell^+_{k+1} - \ell^+_{k-1}}{e^{\beta_0 + (k-1)\beta_r}} + s_{0} & j\text{ even}.
\end{cases}
\end{equation}

\newpage\newpage

We now specialize to the case of hyperbolic laser trajectories, as in \cref{sec_2photon}.  The lasers' initial rapidity is $\kappa = 0$ at coordinate time $t=0$, their initial positions are $\pm L/2$, and their proper accelerations are $\alpha_\pm$.  We suppose that pulse $j$ is fired from its source laser at that laser's proper time $\tau_0 + (j-1)\Delta\tau$.  The initial proper time intervals $s_0, s_1$ become 
\begin{align}
s_0 = \, & e^{\beta_0} \left(\tfrac{L}{2} + \tfrac{1}{\alpha_-} - \tfrac{1}{\alpha_-}e^{-\alpha_- \tau_0}  -r^-_0\right), \\
s_1 = \, & e^{-\beta_0-\beta_r}\left( \tfrac{L}{2} - \tfrac{1}{\alpha_+} + \tfrac{1}{\alpha_+}e^{\alpha_+ (\tau_0+\Delta\tau)} - r^+_0\right) - e^{-\beta_r} s_0.
\end{align}
We have also
\begin{equation}
\ell^\pm_{k+1}-\ell^\pm_{k-1} = \pm \frac{1}{\alpha_\pm} e^{\pm\alpha_\pm \left(\tau_0 + k\Delta\tau\right)} \left(1 - e^{\mp 2\alpha_\pm \Delta\tau}\right),
\end{equation}
for $k\geq 2$, from which we see that all sums in \cref{eq_lmt_s_sol} are geometric.  Then \cref{eq_lmt_s_sol} becomes, for odd $j$,
\begin{widetext}
\begin{multline}
\label{eq_s_wide}
s_j = s_1 + \frac{\left(e^{2 \alpha_+ \Delta \tau }-1\right) \left(e^{\alpha_+ \Delta \tau (j-1)}-e^{(j-1) \beta_r}\right) e^{\alpha_+ \left(\Delta \tau +\tau_0\right)-\beta_0 - j \beta_r}}{\alpha_+ \left(e^{2 \alpha_+ \Delta \tau }-e^{2 \beta_r}\right)} \\
+\frac{\left(e^{2 \alpha_- \Delta \tau }-1\right) \left(e^{\alpha_- \Delta \tau  (j-1)+2 \beta_r}-e^{(j+1) \beta_r}\right) e^{\beta_0-\alpha_- \left(\Delta \tau (j-1)+\tau_0\right)}}{\alpha_- \left(e^{2 \beta_r}-e^{2 \alpha_- \Delta\tau }\right)},
\end{multline}
\end{widetext}
with a similar expression for even $j$.  By virtue of \cref{eq_rp_lmt,eq_rm_lmt}, all atom coordinates $r^\pm_j$ can be determined immediately from $s_j$.  
The contribution of this LMT sequence to clock phase is given by a sum of $s_j$ for $j$ odd times $\omega_a$, and the contribution to propagation phase is given by a sum of all $s_j$ times $m$.  From \cref{eq_s_wide} we see that these phase contributions are combinations of geometric sums, which can thus again be evaluated explicitly.  

We now sketch the remaining steps in computing the full LMT phase shift, providing details in \cite{math_supp}.  The interferometer of \cref{fig_lmt} consists of two LMT sequences for each arm, the first beginning with a pulse from the lower laser and the second beginning with a pulse from the upper laser.  The final atom position computed from the first LMT pulse sequence provides the initial condition for the second.  The clock and propagation phases are sums of proper time intervals $s_j$ for each LMT segment.  To compute the separation phase, one must propagate the upper arm from the central $\pi$ pulse to the final beamsplitter $\pi/2$ pulse and compute the difference of its position and that of the lower arm after the final LMT sequence (this step is straightforward as the upper arm follows a straight line trajectory). 

As in the single-direction case, the laser phase requires additional problem data.  Unlike the single-direction case, there does not appear to be a natural analog of a resonantly chirped laser phase.  This is due to the fact that maintaining the resonance condition for an LMT \ai requires the laser frequency in the lab frame to shift between subsequent pulses, and there is not a unique way to accomplish this. We discuss this issue in \cref{app_lmt_laser_phase}.  For now, we simply assume zero laser phase.  The full expression for the phase shift of an LMT sequence of order $N$ is then the following:
\begin{widetext}
\begin{align} 
\Delta\phi = \, & m \sinh{\beta_r}\Big[\tfrac{\sinh{(\alpha_- T/2)}}{\alpha_-/2} e^{\beta_0+N\beta_r-\alpha_- T}  \left( e^{\alpha_-(\frac{T}{2}-(N-1)\Delta\tau)} S(\beta_r - \alpha_-\Delta\tau)-e^{-\alpha_-(\frac{T}{2}-(N-1)\Delta\tau)} S(\beta_r + \alpha_-\Delta\tau)\right) \nonumber\\ 
& + \tfrac{\sinh{(\alpha_+ T/2)}}{\alpha_+/2} e^{-\beta_0-N\beta_r+\alpha_+ T}  \left( e^{\alpha_+(\frac{T}{2}-N\Delta\tau)} S(\beta_r + \alpha_+\Delta\tau)-e^{-\alpha_+(\frac{T}{2}-N\Delta\tau)} S(\beta_r - \alpha_+\Delta\tau)\right)\Big] \label{eq_lmt_exact_wide}
\end{align}
\end{widetext}
where $S(x):= \frac{\sinh{N x}}{\sinh{x}}$ and $\beta_r$ is that of \cref{eq_ebr}. 
We remark that reaching the expression in \cref{eq_lmt_exact_wide} requires significant cancellation and simplification among the propagation, clock, and separation phases.  The existence of such a comparatively simple final formula may be an indication of a deep principle which is not immediately evident in our formalism for computing \ai phase shifts. 

\section{Off-Shell Transitions}
\label{sec_off_shell}
We have so far restricted to cases where atom optics transitions are perfectly on-shell, so that after the atom-light interaction the norm of the atom's \mom as determined by atomic structure coincides with that implied by \mom conservation using the nominal photon \moma of the atom optics pulses. 
Our focus on this case is partly because, to the best of our knowledge, an authoritative answer from experiment or first-principles theory as to the relativistically correct prescription for the boost imparted to an atom by an off-shell laser interaction is not known.  However, given such a prescription, it is straightforward to adapt the above results to compute phase shifts for off-shell pulses, as we shall show presently.  Additionally, we will see that the quantitative differences between candidate prescriptions are typically quite small, such that this question may be of more academic than practical interest. 

We emphasize that the correct relativistic prescription for computing \ai boost angles is in principle a straightforward scattering problem, and similar effects have been considered previously using perturbative frameworks \cite{Werner2024}.
However, we are not aware of any fully covariant analysis which accounts for effects like the finite pulse duration, which are important in this situation.  Deriving the relativistically correct boost angles is beyond the scope of the present work, but we will analyze three prescriptions for computing off-shell atom boosts which are plausible a priori and see how some of them exhibit untenable features.  For clarity, we will refer to these three prescriptions as the ``energy-momentum conservation'', ``spatial momentum conservation'', and ``frequency selectivity'' prescriptions.  In all prescriptions we ignore finite wavepacket spread in position and momentum, which is consistent with the short pulse limit.  After discussing the relative merits of these prescriptions we will analyze off-shell \ai sequences for both the clock and two-photon cases in subsections below, using the frequency selectivity prescription. 

The energy-momentum conservation prescription determines the boost imparted to the atom by assuming that the atom's final energy-momentum is precisely its initial energy-momentum plus the nominal energy-momenta of any absorbed photons, minus the nominal energy-momenta of emitted photons.  The defect of this prescription is that it does enforce the correct norm for the atom's energy-momentum. In the ground state, the norm is given by the atomic mass $m$, and in the excited state it is $m+\omega_a$. For short-lived excited states, this norm has some imprecision due to the finite lifetime of the state.  However, the norm dictated by the energy-momentum conservation prescription can generally have a much wider range given by the inverse of the light pulse duration.

The spatial momentum conservation prescription supposes that the atom-light interaction conserves spatial momentum but not necessarily energy.  This prescription is essentially that employed in \cite{GRAI}, and computes the final velocity of the atom by assuming that in the rest frame of the atom just before the atom-light interaction the atom's final spatial momentum is given by the net spatial momentum of the atom optics photons, while the time component of the final atom energy-momentum is fixed by demanding that the energy-momentum have the norm prescribed by atomic structure.  This prescription is equivalent to computing the atom-optics dynamics using non-relativistic quantum mechanics in the incoming atom local Lorentz frame.  While this satisfies covariance, it can be shown that it has the following defect: If we compute the boost $\beta_{g\rightarrow e}$ imparted to an atom by a $\pi$ pulse which causes the atom to transition from its ground state to an excited state and the boost $\beta_{e\rightarrow g}$ from an identical $\pi$ pulse which causes the atom to transition from the excited state to the ground state, it is found that $\beta_{g\rightarrow e} \neq -\beta_{e\rightarrow g}$.  This is physically unreasonable, since two $\pi$ pulses should return the atom to its initial \mom state.  Moreover, this prescription can result in different phases for the two output ports (arising from a difference in separation phase; cf. the discussion after \cref{eq_clock_sep}), which is forbidden by unitarity. 

The frequency selectivity prescription is based on the observation that any realistic atom optics pulse will have finite duration, and therefore non-zero frequency bandwidth.  This prescription supposes that the frequency component which effectively drives the atom transition is always precisely resonant.  This prescription is covariant and ensures that two subsequent $\pi$ pulses return the atom to its initial \mom state.  There may be additional laser phase contributions due to finite pulse duration effects, which are beyond the scope of this work. 
We note however that since it does not affect the atom kinematics its contribution to a total \ai phase shift can be incorporated a posteriori.

Note that the fractional differences between the boost angles predicted by each of these prescriptions is normally quite small---on the order of the ratio of the laser detuning from resonance (typically kHz) to the atomic transition frequency (typically hundreds of THz).  Moreover, it is possible that a full treatment of the relevant scattering processes might include additional effects such as wavepacket distortion, which go beyond the short pulse approximation for treating \ai dynamics. 

\subsection{Off-Shell Clock Atom Interferometers}
Under the frequency selectivity prescription, all boosts in a clock atom interferometer are the same as in the perfectly resonant case, and so the kinematic contributions to the phase shift are identical to those computed above, e.g. \cref{eq_kin1}.  The only difference to the full phase shift in the present case is the laser phase, which is now less constrained than before.  (Note that while off-shell transitions have minimal effect on the phase shift, they may have a significant effect on \ai contrast.)  A new case of considerable practical interest we can now consider is that in which the laser frequencies are chirped at a generic rate.  

Consider a single-direction Mach-Zehnder clock \ai with the laser frequency chirped according to 
\begin{align}
\omega(\tau) & = \omega_\ell e^{-\gamma \tau}, \\
\phi(\tau) & = \phi(0) -\frac{\omega_\ell}{\gamma} \left(e^{-\gamma \tau}-1\right)
\end{align}
Then from \cref{eq_laser_phase} the laser phase becomes 
\begin{equation}
\Delta\phi_{\text{laser}} = -\frac{\omega_\ell}{\gamma} \left(e^{-\gamma T} -1\right)^2.
\end{equation}

The full phase shift for a Mach-Zehnder clock \ai with laser acceleration $g$ is then 
\begin{equation}
\Delta\phi = e^{\beta_0} \omega_a \left(1+\frac{\omega_a}{2m}\right)\frac{\left(e^{-g T} - 1\right)^2}{g} - \omega_\ell \frac{\left(e^{-\gamma T} -1\right)^2}{\gamma}.
\end{equation}
To recover the resonant case considered previously, we set the laser frequency to be resonant $\omega_\ell\rightarrow e^{\beta_0}\omega_a \left(1+\frac{\omega_a}{2m}\right)$ and we let the chirp rate coincide with the laser acceleration $\gamma\rightarrow g$.  

Note that the same considerations apply to LMT sequences: The kinematic phases are exactly the same as in the on-shell case (cf. \cref{sec_lmt}), and only the laser phase and possibly the contrast of the \ai will be affected by letting the pulses be off-shell.  The laser phase in this case is described in \cref{app_lmt_laser_phase}. 

\subsection{Off-Shell Raman and Bragg Atom Interferometers}
\label{sec_off_shell_raman}

\begin{table*}[t]
\caption{\label{table_raman} Leading order terms in each of the variables $k_\text{eff}$, $\Delta\omega_a$, $v_0$, $H$, and $L$ for the phase shift $\Delta\phi$ of a Raman \ai in the case of zero laser chirp and in the case of near resonant laser chirp.  We use $k_\text{eff}$ here instead of $\omega_{\pm,0}$ to facilitate comparisons with other literature, e.g. \cite{GRAI,Werner2024,niehof2025finite}; see \cref{eq_keff} for the relationship between these parameters. For each of these variables, we list the term in a series expansion of $\Delta\phi$ which contains that variable and which has lowest total order with respect to all variables as well as $T$; in some cases there are multiple such terms depending on either $k_\text{eff}$ or $\Delta\omega_a$, and in these cases we show only the $k_\text{eff}$ term.  Note that there are many terms not shown which for typical experimental parameter values are larger in magnitude than some of the terms shown here.}
\begin{ruledtabular}
\begin{tabular}{c c c c c c c c c c c c}
& & & & \multicolumn{4}{c}{\textbf{Zero Chirp, $\gamma_\pm=0$}} & & & &  \\[1ex]
$\frac{\alpha_+ + \alpha_-}{2}k_\text{eff} T^2$ & $+$ &
$\frac{\alpha_- - \alpha_+}{2} \Delta \omega_a T^2$ & $+$ &
$\frac{\alpha_- + 3\alpha_+}{2} \, v_0 k_\text{eff} T^2$ & $-$ &
$\alpha_+\left(\alpha_- \! + \! 2\alpha_+\right) H k_\text{eff} T^2$ & $-$ &
$\frac{1}{2}\alpha_+ \left(\alpha_- \! + \! 2\alpha_+\right) L v_0 k_\text{eff} T^2$ & $+$ & 
$\cdots$ \\[3ex]

& & & & \multicolumn{4}{c}{\textbf{Near-Resonant Chirp, $\gamma_\pm=\pm\alpha_\pm \pm \delta\gamma_\pm$}} & & & &  \\[1ex]
$-\frac{\delta\gamma_+ + \delta\gamma_-}{2}k_\text{eff} T^2$ & $+$ &
$\frac{\delta\gamma_+ - \delta\gamma_-}{2} \Delta \omega_a T^2$ & $-$ &
$\frac{\delta\gamma_- + 3\delta\gamma_+}{2} \, v_0 k_\text{eff} T^2$ &  $+$ &
$\delta\gamma_+\left(\alpha_- \! + \! \alpha_+\right) H k_\text{eff} T^2$ & $+$ &
$\frac{1}{2}\delta\gamma_+ \left(\alpha_- \! + \! \alpha_+\right) L v_0 k_\text{eff} T^2$ & $+$ & 
$\cdots$ \\[3ex]
\end{tabular}
\end{ruledtabular}
\end{table*}

With the frequency selectivity prescription for boost angles, an off-shell Raman or Bragg \ai admits an exact solution, but the resulting expressions are sufficiently complicated that we do not present them here in full.  In this section we sketch how the calculation proceeds, discuss some of the interesting terms that arise, and provide full details in \cite{math_supp}.  We also provide a full expression for a special case of a Bragg interferometer in \cref{app_bragg}. 

We consider a Raman or Bragg Mach-Zehnder \ai where the upper and lower lasers follow hyperbolic trajectories as before and are chirped at respective rates $\gamma_+$ and $\gamma_-$, which are allowed to differ from $\pm\alpha_\pm$.  We suppose that the upper laser is passive and always on, while the lower laser is the control laser and thus is pulsed and determines the coordinates of all atom-light interaction points.  Since the upper laser has no pulse envelope, we approximate its frequency as being perfectly monochromatic for the purposes of frequency selection.  Thus all frequency selection will apply only to the frequency of the lower laser.  This implies that the boosts imparted to the atom are determined from the frequency of the upper laser in the lab frame,
\begin{equation}
\label{eq_raman_lab_omega}
\omega_+(\tau) e^{-\alpha_+ \tau} = \omega_{+,0} e^{(\gamma_+-\alpha_+) \tau}.
\end{equation}
The kinematics of the lower laser will affect the boost indirectly by determining at which proper time the upper laser interacts with the atom.  

Due to the interdependence of the atom-light interaction coordinates and the boost imparted to the atom, the logical flow for computing the atom trajectory now has additional steps compared to previously considered \ai sequences.  Given an initial atom position and boost, we first propagate the atom to its next light interaction point with fixed coordinate $\ell^-$.  Then the atom's \lcp coordinate $r^+$ at this interaction point determines the frequency of the upper laser for this interaction.  Finally, the boost imparted to the atom can be deduced from this upper laser frequency via \cref{eq_raman_boost_2,eq_raman_boost_2_emit}.  This process is repeated for each pulse from the lower laser. 

We take the atom's initial position at coordinate time $t=0$ to be $x=H$ so that its initial LC coordinates are $r^\pm_0=\pm H$, and its initial boost is $\beta_0$.  The lasers start at position $\pm \frac{L}{2}$ at $t=0$ with initial rapidity zero.  Thus the lower laser trajectory has \lcm coordinate
\begin{equation}
\ell^-(\tau)=\frac{L}{2} + \frac{1}{\alpha_-} - \frac{1}{\alpha_-} e^{-\alpha_- \tau},
\end{equation}
while the upper laser trajectory has \lcp coordinate 
\begin{equation}
\ell^+(\tau)=\frac{L}{2} - \frac{1}{\alpha_+} + \frac{1}{\alpha_+} e^{\alpha_+ \tau}.
\end{equation}
The frequency (in the laser's instantaneous rest frame) and phase of the upper laser can be expressed in terms of $\ell^+(\tau)$,
\begin{align}
\omega_+(\tau) & = \omega_{+,0} e^{\gamma_+\tau} = \omega_{+,0} \left(\alpha_+\ell^+ - \frac{1}{2} \alpha_+ L + 1\right)^{\frac{\gamma_+}{\alpha_+}}, \label{eq_raman_omega_vs_l}\\
\phi_{+}(\tau) & = \frac{\omega_{+,0}}{\gamma_+} e^{\gamma_+\tau} = \frac{\omega_{+,0}}{\gamma_+} \left(\alpha_+\ell^+ - \frac{1}{2} \alpha_+ L + 1\right)^{\frac{\gamma_+}{\alpha_+}}.
\end{align}
In the degenerate case $\gamma_+=0$, the latter expression becomes instead
\begin{equation}
\phi_{+}(\tau) = \omega_{+,0} \tau = \frac{\omega_{+,0}}{\alpha_+} \log\left(\alpha_+\ell^+ - \frac{1}{2} \alpha_+ L + 1\right).
\end{equation}

Taking the initial beamsplitter pulse to occur at the lower laser's proper time $\tau=0$, the first atom-light interaction point will be at 
\begin{equation}
\dvec{r^+_1}{r^-_1} = \dvec{H+e^{2\beta_0}\left(\frac{L}{2}+H\right)}{\frac{L}{2}}.
\end{equation}
From \cref{eq_raman_lab_omega,eq_raman_omega_vs_l}, for this interaction the frequency of the upper laser in the lab frame will be 
\begin{equation}
\label{eq_off_shell_lp_freq}
\omega_{+,1} = \omega_{+,0}\left(\!\alpha_+ H \!\left(e^{2\beta_0} \! + \! 1\right) \! + \! \frac{\alpha_+ L}{2} \left(e^{2\beta_0} \! - \! 1\right) \! + \! 1 \! \right)^{\!\!\frac{\gamma_+ - \alpha_+}{\alpha_+}} \!\!\! ,
\end{equation}
and using \cref{eq_raman_boost_2} 
the boost $\beta_{r,1}$ imparted to the upper arm of the interferometer will satisfy
\begin{equation}
e^{\beta_{r,1}} = \frac{m+\Delta\omega_a}{m - 2e^{\beta_0} \omega_{+,1}}.
\end{equation}
We now must repeat this process for the mirror $\pi$ pulse and final beam splitter.  Note that the ratio $\gamma_+/\alpha_+$ in the exponent of \cref{eq_off_shell_lp_freq} makes the resulting expressions much more complicated than the analogous clock \ai.  Additionally, the variation in boost from different atom optics pulses means that the two wavepackets in a given output port will now no longer have exactly equal \moma, i.e. there will be phase shear on the output port.  

In the special case where $\Delta\omega_a=0$ (i.e. a Bragg interferometer), $\gamma_\pm=0$, and the atom initial conditions are $\beta_0=0$, $H=0$, the full phase shift is simple enough that we present it in \cref{app_bragg}.  The fully general case is in \cite{math_supp}.  We comment now on some interesting features of this solution.  

\cref{table_raman} shows the lowest order terms in a series expansion of the full Raman phase shift containing each of the variables $k_\text{eff}$, $\Delta\omega_a$, $v_0$, $H$, and $L$.  (We use the effective wavenumber $k_\text{eff}:=\omega_{+,0}+\omega_{-,0}$ rather than $\omega_{+,0}$ to facilitate comparison to previous work; cf. \cref{eq_keff}.)  We see that the term $\frac{\alpha_+ + \alpha_-}{2}k_\text{eff} T^2$ in the zero chirp case recovers the classical phase shift when $\alpha_+ = g$ and $\alpha_- = g$.  In the case of near-resonant chirp, the fact that all terms contain a factor of $\delta\gamma_\pm := \pm \gamma_\pm -\alpha_\pm$ recovers the previous result that the phase shift vanishes for a perfectly resonant chirp. 

The dependence of $\Delta\phi$ on parameters such as the atom initial position $H$ is an effect which appears not to have been emphasized before, though it is analogous to previous work on consequences of laser propagation delay \cite{le2007influence,cheng2015influence,niehof2025finite}.
The origin of this effect can be traced to the implicit dependence on the atom's position of the frequency of the upper laser at the proper time when it emits a pulse that will interact with the atom. 

\vspace{-0.3cm}
\section{Discussion}
\label{sec_discussion}
\vspace{-0.2cm}

In this section we reflect on some general features of the atom interferometers we have analyzed above, emphasizing what conditions are needed for the interferometer to admit an exact analytical solution and what can be said in situations where these conditions are violated.  We informally identify four key ingredients for solvability: affine linearity of atom and laser trajectories; analytic expressions for recoil boosts; analytic expressions for laser pulse coordinates; and, for obtaining general solutions in cases like LMT with a parametric number of pulses, analytic summability of recoil boosts and laser pulse coordinates.

The fact that flat spacetime has a coordinate system in which atom trajectories are affine linear ensures that atom trajectories can be fully solved for arbitrary pulse sequences provided the boost angles at atom-light interactions are known.  However, this is clearly a sufficient but not necessary condition for solvability, as one may simply transform the exact solutions we have found above into an arbitrary coordinate system where the trajectories are fully non-linear.  Nevertheless, in typical curved spacetimes, such as Schwarzschild space, deriving exact solutions appears challenging.  In such cases, exact flat space solutions may still prove useful as the basis for perturbative treatments.  Beginning with a zeroth order solution which already includes all special relativistic phenomenology may facilitate the understanding of effects due to spacetime curvature.  Moreover, we note that it is possible to define LC coordinates with respect to two arbitrary laser trajectories even in curved spacetime, which may have computational benefits analogous to those exploited systematically in the present work.  

The relativistically correct prescription for recoil boosts of off-shell transitions is the largest open question posed by our work.  While we have performed off-shell \ai calculations using the frequency selectivity prescription, we emphasize that we do not claim to have given rigorous justification that this prescription is correct.  It bears repeating that given any similar prescription which provides analytical expressions for the recoil boost angle of all atom-light interactions, exact solution of all properties of the \ai in flat spacetime should be possible.  However, as the case of the off-shell Raman \ai shows, analytic solutions can be quite complex for complicated recoil boost prescriptions.  The frequency selectivity prescription has the great virtue of providing essentially the simplest possible off-shell behavior for clock \ais, since all momentum states of the atom lie on a uniform ladder of boost angles.  This property is not shared even by other simple prescriptions we have considered (e.g. spatial momentum prescription), and may to some extent justify using this prescription to understand the structure of relativistic \ais even if it proves to only be an approximation.

The necessity for laser pulse coordinates to have analytic expressions for exact solvability is evident from formulas such as \cref{eq_kin1}.  In addition to the laser trajectories we have considered above, another case of practical interest for which \ai trajectories should be exactly solvable is that in which the lasers are in uniform rotational motion.  This is especially relevant for understanding the effect of the Coriolis force on \ai phase shifts \cite{dubetsky2006atom,lan2012influence}.  Note that our analysis above used a 1+1 dimensional spacetime, whereas describing rotating lasers requires a 3+1 dimensional spacetime, as well as additional problem data concerning the transverse structure of atom optics pulses.  In particular, the same light pulse null geodesics will generally not interact with both arms of the interferometer in this case, meaning that the data for the pulse must include a transverse phase profile and an initial space-like surface marking the beginning of the pulse. 

The fact that a clock LMT \ai admits an exact solution for general order $N$ requires much stronger mathematical structure than the existence of solutions for particular orders.  We have argued above that any particular \ai trajectory is solvable in flat space provided the boost angles can be expressed analytically.  However, for the general LMT case to be solvable, we needed the fact that the boost angles of successive segments during an LMT sequence were in geometric progression.  This notably would fail to hold in the off-shell case if the frequency selectivity prescription for recoil boost were not employed. 

As a final remark, we note that many of the sums encountered throughout the present work can be interpreted as generating functions of laser pulse coordinates.  For instance, for a single-direction \ai, the record $A$ takes the form 
\begin{equation}
A = -\sum_j z^j \left(\ell^-_{j+1}-\ell^-_j\right)
\end{equation}
evaluated at $z=-1$.  Similarly, the sums appearing in the expression \cref{eq_lmt_s_sol} for proper time interval $s_j$ of an LMT sequence take the form 
\begin{align}
\mathcal{L}^-(z) & = \sum\limits_{\substack{k=2 \\ k\text{ even}}}^{2N-2} z^k \left(\ell^-_{k+1}-\ell^-_{k-1}\right),\\
\mathcal{L}^+(z) & = \sum\limits_{\substack{k=1 \\ k\text{ odd}}}^{2N-1} z^k \left(\ell^+_{k+1}-\ell^+_{k-1}\right),
\end{align}
evaluated at $z=e^{\pm\beta_r}$.  All of these sums have the form of ordinary generating functions associated to series of laser pulse coordinate differences.  Moreover, it can be shown that the clock and propagation phases in the LMT case can be expressed in terms of derivatives of the generating functions $\mathcal{L}^\pm(z)$.  It may be possible to systematically exploit properties of such generating functions for analysis of complicated \ais, such as resonant LMT sequences \cite{graham2016resonant}, which are likely to be very complex to analyze by a direct approach. 

\vspace{-0.2cm}
\section{Conclusion}
\label{sec_conclusion}
\vspace{-0.2cm}
We have derived relativistically exact expressions for a variety of experimentally relevant atom interferometer pulse sequences, including Mach-Zehnder, resonant, and LMT geometries in flat spacetime.  For the case of Mach-Zehnder interferometers, we have provided detailed analyses for both clock atom interferometers as well as Raman or Bragg interferometers.  For on-shell interferometers with all pulses from a single direction, we can completely characterize the phase shift by a sum of laser light cone coordinates at the atom optic emission times.  For the LMT clock atom accelerometer, we have given a general expression for arbitrary LMT order $N$.  In the case of an off-shell Raman or Bragg interferometer, we have demonstrated subtle dependence of the phase shift on parameters such as the atom initial position, which to our knowledge has not been noticed before.  These results and the methods by which they are obtained provide new insights into the relativistic sensitivity of atom interferometry.

\vspace{-0.4cm}
\begin{acknowledgments}
\vspace{-0.2cm}
This work was supported by the National Science Foundation QLCI Award No. OMA-2016244 and the Gordon and Betty Moore Foundation, grant DOI 10.37807/GBMF7945.01.  The authors acknowledge use of Gemini Pro 3.1 and OpenAI Codex for code generation and error checking of code and mathematical formulas. 
\end{acknowledgments}

\appendix

\section{Two-Photon Recoil Boosts}
\label{app_boosts}
We present here analogous formulas to \cref{sec_boosts} for the case of a two-photon Raman or Bragg transition.  We consider two lasers, one below and one above the atom, with respective frequencies $\omega_-$ and $\omega_+$ in the lab frame, or $e^{-\beta_0}\omega_-$ and $e^{\beta_0}\omega_+$ in the atom's rest frame.  If the transition is on-shell and the atom absorbs a photon from the lower laser and emits along the upper laser, its final \mom is $\invec{m+e^{-\beta_0}\omega_- - e^{\beta_0}\omega_+}{e^{-\beta_0}\omega_- + e^{\beta_0}\omega_+}$ in its initial rest frame.  The on-shell condition now requires that the atom's final \mom have norm equal to $m+\Delta\omega_a$, with $\Delta \omega_a$ equal to the ground state splitting frequency in the Raman case and $\Delta\omega_a=0$ in the Bragg case.  This constrains the frequencies $\omega_-, \omega_+$ by 
\begin{equation}
\label{eq_on_shell1}
(m+\Delta\omega_a)^2 = (m+2e^{-\beta_0}\omega_-)(m-2e^{\beta_0}\omega_+),
\end{equation}
or equivalently
\begin{align}
\label{eq_on_shell2}
e^{\beta_0}\omega_+ & = \frac{2 m e^{-\beta_0}\omega_- - 2 m \Delta \omega_a - \Delta \omega_a^2}{2 m+4 e^{-\beta_0}\omega_-}, \\
\label{eq_on_shell3}
e^{-\beta_0}\omega_- & = \frac{2 m e^{\beta_0}\omega_+ + 2 m \Delta \omega_a + \Delta \omega_a^2}{2 m - 4 e^{\beta_0}\omega_+}.
\end{align}
The resulting boost angle $\beta_r$ is then 
\begin{align}
\label{eq_os_raman_boost}
\beta_r & = \tanh^{-1}\left( \frac{e^{-\beta_0}\omega_- + e^{\beta_0}\omega_+}{m + e^{-\beta_0}\omega_- - e^{\beta_0}\omega_+} \right).
\end{align}
It is straightforward to show that this satisfies 
\begin{align}
e^{\beta_r} = \frac{m + \Delta\omega_a}{m - 2e^{\beta_0}\omega_+} = \frac{m + 2e^{-\beta_0}\omega_-}{m+\Delta\omega_a}. \label{eq_raman_boost_2}
\end{align}
Similarly, for the reverse process where the atom absorbs from the upper laser and emits into the lower laser, transitioning from a state of mass $m + \Delta\omega_a$ to a state of mass $m$, we have
\begin{align}
e^{\beta_r} = \frac{m}{m + \Delta\omega_a + 2e^{\beta_0}\omega_+} = \frac{m + \Delta\omega_a - 2e^{-\beta_0}\omega_-}{m}. \label{eq_raman_boost_2_emit}
\end{align}

Previous literature on relativistic atom interferometers often gives expressions for Raman or Bragg interferometer phase shifts in terms of the effective wavenumber $k_\text{eff}=\omega_+ + \omega_-$ of the transition.  For an on-shell transition, both laser frequencies can be expressed in terms of $k_\text{eff}$, 

\begin{align}
\omega_\pm & = \frac{k_\text{eff}}{2} \pm \frac{m}{2} \cosh(\beta_0) \mp \nonumber \\ 
& \hspace{-0.4cm} \frac{1}{2}\sqrt{m^2\cosh^2\beta_0 \! + \! 2 m k_\text{eff} \sinh\beta_0 \! + \! k_\text{eff}^2 \! + \! 2 m \Delta \omega_a \! + \! \Delta\omega_a^2} \nonumber \\
& = \frac{1}{2}\left(k_\text{eff} \mp \Delta\omega_a\right) \mp \frac{1}{2}\beta_0 k_\text{eff} + \cdots. \label{eq_keff}
\end{align}

\begin{figure}[b]
\centering
    \includegraphics[width=0.99\linewidth]{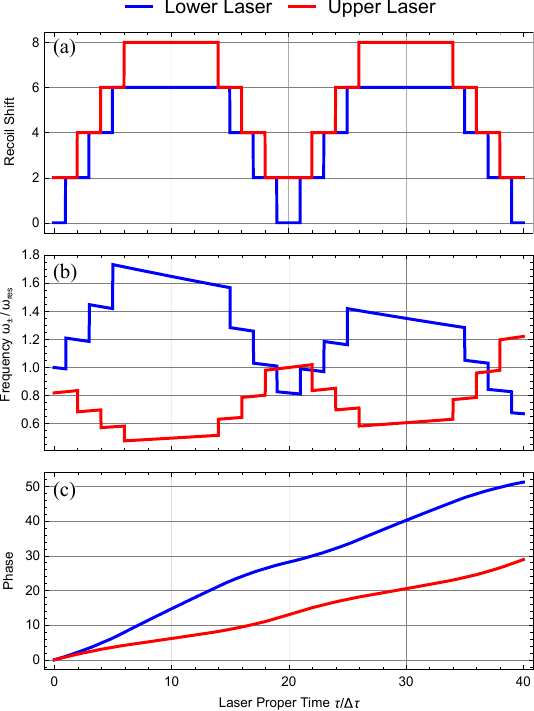}\vspace{-0.38cm}
    \caption{An example frequency and phase schedule for lasers driving an on-shell LMT pulse sequence.  Shown is the case of LMT order $N=4$, $T=20\Delta\tau$, and with frequency jumps occurring at the midpoint between successive pulses.  The laser frequencies must vary in time to cancel Doppler shifts due to both laser acceleration and the recoil of the atom interferometer arms.  \textbf{(a)}  Integer number of photon recoils by which each laser is detuned.  The value $j_\pm$ of such a curve at a given proper time $\tau$ is related to the corresponding laser frequency $\omega_\pm$ by $\omega_\pm(\tau)=\omega_\text{res} e^{\mp\beta_0\mp j \beta_r\pm\alpha_{\pm}\tau}$. \textbf{(b)}  Laser frequencies, including Doppler shifts from both atom recoil and laser acceleration.  \textbf{(c)}  Laser phase, determined by integration of the curves in (b).}
\label{fig_lmt_laser_freq} 
\end{figure}

For phase shift calculations we find it much more practical to work with $\omega_\pm$ than $k_\text{eff}$, and we will thus only employ \cref{eq_keff} in constructing term lists for comparison to other literature. Note that for purposes of series expansion, the boost angle $\beta_0$ is equal to the corresponding coordinate velocity $v_0$ to leading order. 

\section{LMT Laser Phase}
\label{app_lmt_laser_phase}

In this section we survey possible laser frequency and phase schedules for maintaining LMT atom optics perfectly on-shell.  Unlike in the case of a single-direction clock \ai, there is not a unique, natural way to maintain each pulse on-shell.  In the lab frame, the frequency of e.g. the lower laser for pulse $j$ of the initial LMT beamsplitter should satisfy 
\begin{equation}
\omega_{-,j}=e^{\beta_0+(j-1)\beta_r}\omega_\text{res}.
\end{equation}
Practically, experiments involving LMT sequences will discontinuously jump the laser frequency between subsequent pulses to meet this condition, but the specific manner in which this is carried out will affect the resulting atom interferometer phase shift.  For instance, electing to jump the frequency immediately after each pulse, versus symmetrically in between pulses, results in a different laser phase. 

For illustration, we evaluate the laser phase analytically for a particular choice of frequency and phase schedule.  Namely, we take each laser's frequency to jump discontinuously at the proper time midpoint between subsequent pulses, and to be continuously chirped so as to cancel the effect of laser acceleration.  The frequency and phase schedules are shown for the case $N=4$ in \cref{fig_lmt_laser_freq}. 

The laser phase contribution to the \ai phase shift can be shown in this case to be
\begin{equation}
\label{eq_lmt_res_laser}
\Delta\phi_\text{laser} = \Delta\phi_\text{laser}^{(-)} + \Delta\phi_\text{laser}^{(+)} 
\end{equation}
with the contribution from each laser given by
\begin{widetext}
\begin{align}
    \Delta\phi_\text{laser}^{(-)} &= -\omega_\text{res} e^{-\beta_r}\tfrac{\sinh{(\alpha_- T/2)}}{\alpha_-/2} e^{\beta_0 + N\beta_r-\alpha_- T}  \Big[ e^{\alpha_-(\frac{T}{2}-(N-1)\Delta\tau)} \big(S(\beta_r - \alpha_-\Delta\tau) + \sinh\beta_r \,U_-(\beta_r - \alpha_-\Delta\tau) \big) \nonumber\\
    & \hspace{60mm} -e^{-\alpha_-(\frac{T}{2}-(N-1)\Delta\tau)} \big(S(\beta_r + \alpha_-\Delta\tau) + \sinh\beta_r \, U_-(\beta_r + \alpha_-\Delta\tau) \big) \Big]\nonumber\\ 
    \Delta\phi_\text{laser}^{(+)} &= -\omega_\text{res} e^{-\beta_r} \tfrac{\sinh{(\alpha_+ T/2)}}{\alpha_+/2} e^{-\beta_0-N\beta_r+\alpha_+ T}  \Big[e^{\alpha_+(\frac{T}{2}-N\Delta\tau)} \big(S(\beta_r + \alpha_+\Delta\tau) + \sinh\beta_r \, U_+(\beta_r + \alpha_+\Delta\tau) \big)\nonumber\\
    &\hspace{60mm} -e^{-\alpha_+(\frac{T}{2}-N\Delta\tau)} \big(S(\beta_r - \alpha_+\Delta\tau) + \sinh\beta_r \, U_+(\beta_r - \alpha_+\Delta\tau)\big) \Big], \nonumber 
\end{align}
\end{widetext}
where $S(x):=\frac{\sinh Nx}{\sinh x}$ and $U_\pm(x) \equiv \frac{N e^{\mp (N-1) x} - S(x)}{\sinh{x}}$.  We have written this equation in a manner emphasizing similarity with \cref{eq_lmt_exact_wide}.  Note that $\omega_\text{res} = m e^{\beta_r} \sinh \beta_r$, so the kinematic phase contributions of \cref{eq_lmt_exact_wide} cancel all but the terms involving $U_\pm$ in this laser phase, which are relativistically suppressed compared to either the full kinematic or full laser phase.  

In the case of off-shell laser pulses, the laser phase for a generic starting frequency $\omega_{\pm,0}$ and chirp rate $\gamma_\pm$ can be recovered from the preceding expressions via the substitutions $\omega_\text{res}\rightarrow \omega_{\pm,0}$, $\alpha_\pm\rightarrow\pm\gamma_\pm$, and $\beta_0\rightarrow 0$. 

\newpage 
\section{Phase Shift for Bragg Interferometer with No Frequency Chirp}
\label{app_bragg}

We state here the full phase shift $\Delta\phi$ for a Bragg \ai for the special case when both lasers have no frequency chirp and the atom has initial position and spatial velocity zero.  The logarithmic terms arise from the laser phase of the upper laser, while all other terms originate from the propagation and separation phases. 

\begin{widetext}
\begin{multline}
\Delta\phi = \\
-\frac{2 m \omega_+^2 e^{-2 \alpha_- T} \left(e^{\alpha_- T}-1\right)^2}{\alpha_- \left(m-2 \omega_+\right)^2}
\frac{\alpha_+ m+e^{\alpha_- T} \left(\alpha_- \left(m-2 \omega_+\right)+\alpha_+ m\right)}{-\alpha_+
   m + e^{\alpha_- T} \left(\alpha_- \left(m-2 \omega_+\right)+\alpha_+ m\right)} \left(\alpha_+ m^2 \left(e^{\alpha_-
   T}-1\right)+\alpha_- \left(m-2 \omega_+\right)^2 e^{\alpha_- T}\right)^{-1}
\\
    \Bigg[\alpha_+^3 m^3 \omega_+ \left(e^{\alpha_-
   T}-1\right)^3 \left(e^{\alpha_- T}+1\right)+\alpha_- \alpha_+^2 m^2 \left(m-2
   \omega_+\right) e^{\alpha_- T} \left(e^{\alpha_- T}-1\right)^2 \left(m
   e^{\alpha_- T}-2 \omega_+\right)
\\
   + \alpha_-^3 \left(m-2 \omega_+\right)^4 e^{4
   \alpha_- T}+\alpha_-^2 \alpha_+ m \left(m-2 \omega_+\right)^2 e^{2 \alpha_-
   T} \left(e^{\alpha_- T}-1\right) \left(\left(2 m-3 \omega_+\right) e^{\alpha_-
   T}+\omega_+\right)\Bigg] 
\\
   \left(\alpha_+^2 m^2-\alpha_+ \left(\alpha_-+\alpha_+\right) m^2
   e^{\alpha_- T}-\alpha_+ m e^{2 \alpha_- T} \left(\alpha_- \left(m-4 \omega_+\right)+\alpha_+ m\right)+e^{3 \alpha_- T} \left(\alpha_- \left(m-2 \omega_+\right)+\alpha_+ m\right)^2\right)^{-1} 
\\
   + \frac{\omega_+}{\alpha_+} \log \left({\scriptstyle
   \frac{\left(\alpha_+ m-e^{\alpha_- T} \left(\alpha_-
   \left(m-2 \omega_+\right)+\alpha_+ m\right)\right)^2 \left(\alpha_+ m^2
   \left(e^{\alpha_- T}-1\right)+\alpha_- \left(m-2 \omega_+\right)^2 e^{\alpha_-
   T}\right)}{\alpha_- \left(m-2 \omega_+\right)^2 \left(\alpha_+^2 m^2-\alpha_+
   \left(\alpha_-+\alpha_+\right) m^2 e^{\alpha_- T}-\alpha_+ m e^{2 \alpha_- T}
   \left(\alpha_- \left(m-4 \omega_+\right)+\alpha_+ m\right)+e^{3 \alpha_- T}
   \left(\alpha_- \left(m-2 \omega_+\right)+\alpha_+
   m\right)^2\right)}}\right)
\end{multline}
\end{widetext}

\bibliography{bibs}

\providecommand{\noopsort}[1]{}\providecommand{\singleletter}[1]{#1}%
\begin{thebibliography}{27}%
\makeatletter
\providecommand \@ifxundefined [1]{%
 \@ifx{#1\undefined}
}%
\providecommand \@ifnum [1]{%
 \ifnum #1\expandafter \@firstoftwo
 \else \expandafter \@secondoftwo
 \fi
}%
\providecommand \@ifx [1]{%
 \ifx #1\expandafter \@firstoftwo
 \else \expandafter \@secondoftwo
 \fi
}%
\providecommand \natexlab [1]{#1}%
\providecommand \enquote  [1]{``#1''}%
\providecommand \bibnamefont  [1]{#1}%
\providecommand \bibfnamefont [1]{#1}%
\providecommand \citenamefont [1]{#1}%
\providecommand \href@noop [0]{\@secondoftwo}%
\providecommand \href [0]{\begingroup \@sanitize@url \@href}%
\providecommand \@href[1]{\@@startlink{#1}\@@href}%
\providecommand \@@href[1]{\endgroup#1\@@endlink}%
\providecommand \@sanitize@url [0]{\catcode `\\12\catcode `\$12\catcode `\&12\catcode `\#12\catcode `\^12\catcode `\_12\catcode `\%12\relax}%
\providecommand \@@startlink[1]{}%
\providecommand \@@endlink[0]{}%
\providecommand \url  [0]{\begingroup\@sanitize@url \@url }%
\providecommand \@url [1]{\endgroup\@href {#1}{\urlprefix }}%
\providecommand \urlprefix  [0]{URL }%
\providecommand \Eprint [0]{\href }%
\providecommand \doibase [0]{https://doi.org/}%
\providecommand \selectlanguage [0]{\@gobble}%
\providecommand \bibinfo  [0]{\@secondoftwo}%
\providecommand \bibfield  [0]{\@secondoftwo}%
\providecommand \translation [1]{[#1]}%
\providecommand \BibitemOpen [0]{}%
\providecommand \bibitemStop [0]{}%
\providecommand \bibitemNoStop [0]{.\EOS\space}%
\providecommand \EOS [0]{\spacefactor3000\relax}%
\providecommand \BibitemShut  [1]{\csname bibitem#1\endcsname}%
\let\auto@bib@innerbib\@empty
\bibitem [{\citenamefont {Kasevich}\ and\ \citenamefont {Chu}(1992)}]{kasevich1992measurement}%
  \BibitemOpen
  \bibfield  {author} {\bibinfo {author} {\bibfnamefont {M.}~\bibnamefont {Kasevich}}\ and\ \bibinfo {author} {\bibfnamefont {S.}~\bibnamefont {Chu}},\ }\bibfield  {title} {\bibinfo {title} {Measurement of the gravitational acceleration of an atom with a light-pulse atom interferometer},\ }\href {https://doi.org/10.1007/BF00325375} {\bibfield  {journal} {\bibinfo  {journal} {Applied Physics B}\ }\textbf {\bibinfo {volume} {54}},\ \bibinfo {pages} {321} (\bibinfo {year} {1992})}\BibitemShut {NoStop}%
\bibitem [{\citenamefont {Graham}\ \emph {et~al.}(2013)\citenamefont {Graham}, \citenamefont {Hogan}, \citenamefont {Kasevich},\ and\ \citenamefont {Rajendran}}]{Graham2013}%
  \BibitemOpen
  \bibfield  {author} {\bibinfo {author} {\bibfnamefont {P.~W.}\ \bibnamefont {Graham}}, \bibinfo {author} {\bibfnamefont {J.~M.}\ \bibnamefont {Hogan}}, \bibinfo {author} {\bibfnamefont {M.~A.}\ \bibnamefont {Kasevich}},\ and\ \bibinfo {author} {\bibfnamefont {S.}~\bibnamefont {Rajendran}},\ }\bibfield  {title} {\bibinfo {title} {New method for gravitational wave detection with atomic sensors},\ }\href {https://doi.org/10.1103/PhysRevLett.110.171102} {\bibfield  {journal} {\bibinfo  {journal} {Phys. Rev. Lett.}\ }\textbf {\bibinfo {volume} {110}},\ \bibinfo {pages} {171102} (\bibinfo {year} {2013})}\BibitemShut {NoStop}%
\bibitem [{\citenamefont {Dimopoulos}\ \emph {et~al.}(2008{\natexlab{a}})\citenamefont {Dimopoulos}, \citenamefont {Graham}, \citenamefont {Hogan}, \citenamefont {Kasevich},\ and\ \citenamefont {Rajendran}}]{AGIS2008}%
  \BibitemOpen
  \bibfield  {author} {\bibinfo {author} {\bibfnamefont {S.}~\bibnamefont {Dimopoulos}}, \bibinfo {author} {\bibfnamefont {P.~W.}\ \bibnamefont {Graham}}, \bibinfo {author} {\bibfnamefont {J.~M.}\ \bibnamefont {Hogan}}, \bibinfo {author} {\bibfnamefont {M.~A.}\ \bibnamefont {Kasevich}},\ and\ \bibinfo {author} {\bibfnamefont {S.}~\bibnamefont {Rajendran}},\ }\bibfield  {title} {\bibinfo {title} {Atomic gravitational wave interferometric sensor},\ }\href {https://doi.org/10.1103/PhysRevD.78.122002} {\bibfield  {journal} {\bibinfo  {journal} {Phys. Rev. D}\ }\textbf {\bibinfo {volume} {78}},\ \bibinfo {pages} {122002} (\bibinfo {year} {2008}{\natexlab{a}})}\BibitemShut {NoStop}%
\bibitem [{\citenamefont {Dimopoulos}\ \emph {et~al.}(2008{\natexlab{b}})\citenamefont {Dimopoulos}, \citenamefont {Graham}, \citenamefont {Hogan},\ and\ \citenamefont {Kasevich}}]{GRAI}%
  \BibitemOpen
  \bibfield  {author} {\bibinfo {author} {\bibfnamefont {S.}~\bibnamefont {Dimopoulos}}, \bibinfo {author} {\bibfnamefont {P.~W.}\ \bibnamefont {Graham}}, \bibinfo {author} {\bibfnamefont {J.~M.}\ \bibnamefont {Hogan}},\ and\ \bibinfo {author} {\bibfnamefont {M.~A.}\ \bibnamefont {Kasevich}},\ }\bibfield  {title} {\bibinfo {title} {General relativistic effects in atom interferometry},\ }\href {https://doi.org/10.1103/PhysRevD.78.042003} {\bibfield  {journal} {\bibinfo  {journal} {Phys. Rev. D}\ }\textbf {\bibinfo {volume} {78}},\ \bibinfo {pages} {042003} (\bibinfo {year} {2008}{\natexlab{b}})}\BibitemShut {NoStop}%
\bibitem [{\citenamefont {Storey}\ and\ \citenamefont {Cohen-Tannoudji}(1994)}]{storey1994feynman}%
  \BibitemOpen
  \bibfield  {author} {\bibinfo {author} {\bibfnamefont {P.}~\bibnamefont {Storey}}\ and\ \bibinfo {author} {\bibfnamefont {C.}~\bibnamefont {Cohen-Tannoudji}},\ }\bibfield  {title} {\bibinfo {title} {The {F}eynman path integral approach to atomic interferometry. a tutorial},\ }\href {https://doi.org/10.1051/jp2:1994103} {\bibfield  {journal} {\bibinfo  {journal} {J. Phys. II}\ }\textbf {\bibinfo {volume} {4}},\ \bibinfo {pages} {1999} (\bibinfo {year} {1994})}\BibitemShut {NoStop}%
\bibitem [{\citenamefont {Antoine}\ and\ \citenamefont {Bord{\'e}}(2003)}]{antoine2003exact}%
  \BibitemOpen
  \bibfield  {author} {\bibinfo {author} {\bibfnamefont {C.}~\bibnamefont {Antoine}}\ and\ \bibinfo {author} {\bibfnamefont {C.~J.}\ \bibnamefont {Bord{\'e}}},\ }\bibfield  {title} {\bibinfo {title} {Exact phase shifts for atom interferometry},\ }\href {https://doi.org/10.1016/S0375-9601(02)01625-0} {\bibfield  {journal} {\bibinfo  {journal} {Phys. Lett. A}\ }\textbf {\bibinfo {volume} {306}},\ \bibinfo {pages} {277} (\bibinfo {year} {2003})}\BibitemShut {NoStop}%
\bibitem [{\citenamefont {Tan}\ and\ \citenamefont {Shao}(2017)}]{tan2017calculating}%
  \BibitemOpen
  \bibfield  {author} {\bibinfo {author} {\bibfnamefont {Y.-J.}\ \bibnamefont {Tan}}\ and\ \bibinfo {author} {\bibfnamefont {C.-G.}\ \bibnamefont {Shao}},\ }\bibfield  {title} {\bibinfo {title} {Calculating the finite-speed-of-light effect in atom gravimeters with general relativity},\ }in\ \href {https://doi.org/10.1142/9789813148505_0078} {\emph {\bibinfo {booktitle} {Proceedings of the Seventh Meeting on CPT and Lorentz Symmetry}}}\ (\bibinfo {organization} {World Scientific},\ \bibinfo {year} {2017})\ pp.\ \bibinfo {pages} {289--291}\BibitemShut {NoStop}%
\bibitem [{\citenamefont {Roura}(2020)}]{Roura2020}%
  \BibitemOpen
  \bibfield  {author} {\bibinfo {author} {\bibfnamefont {A.}~\bibnamefont {Roura}},\ }\bibfield  {title} {\bibinfo {title} {Gravitational redshift in quantum-clock interferometry},\ }\href {https://doi.org/10.1103/PhysRevX.10.021014} {\bibfield  {journal} {\bibinfo  {journal} {Phys. Rev. X}\ }\textbf {\bibinfo {volume} {10}},\ \bibinfo {pages} {021014} (\bibinfo {year} {2020})}\BibitemShut {NoStop}%
\bibitem [{\citenamefont {Ufrecht}\ and\ \citenamefont {Giese}(2020)}]{ufrecht2020perturbative}%
  \BibitemOpen
  \bibfield  {author} {\bibinfo {author} {\bibfnamefont {C.}~\bibnamefont {Ufrecht}}\ and\ \bibinfo {author} {\bibfnamefont {E.}~\bibnamefont {Giese}},\ }\bibfield  {title} {\bibinfo {title} {Perturbative operator approach to high-precision light-pulse atom interferometry},\ }\href {https://doi.org/10.1103/PhysRevA.101.053615} {\bibfield  {journal} {\bibinfo  {journal} {Phys. Rev. A}\ }\textbf {\bibinfo {volume} {101}},\ \bibinfo {pages} {053615} (\bibinfo {year} {2020})}\BibitemShut {NoStop}%
\bibitem [{\citenamefont {Werner}\ \emph {et~al.}(2024)\citenamefont {Werner}, \citenamefont {Schwartz}, \citenamefont {Kirsten-Siem\ss{}}, \citenamefont {Gaaloul}, \citenamefont {Giulini},\ and\ \citenamefont {Hammerer}}]{Werner2024}%
  \BibitemOpen
  \bibfield  {author} {\bibinfo {author} {\bibfnamefont {M.}~\bibnamefont {Werner}}, \bibinfo {author} {\bibfnamefont {P.~K.}\ \bibnamefont {Schwartz}}, \bibinfo {author} {\bibfnamefont {J.-N.}\ \bibnamefont {Kirsten-Siem\ss{}}}, \bibinfo {author} {\bibfnamefont {N.}~\bibnamefont {Gaaloul}}, \bibinfo {author} {\bibfnamefont {D.}~\bibnamefont {Giulini}},\ and\ \bibinfo {author} {\bibfnamefont {K.}~\bibnamefont {Hammerer}},\ }\bibfield  {title} {\bibinfo {title} {Atom interferometers in weakly curved spacetimes using {B}ragg diffraction and {B}loch oscillations},\ }\href {https://doi.org/10.1103/PhysRevD.109.022008} {\bibfield  {journal} {\bibinfo  {journal} {Phys. Rev. D}\ }\textbf {\bibinfo {volume} {109}},\ \bibinfo {pages} {022008} (\bibinfo {year} {2024})}\BibitemShut {NoStop}%
\bibitem [{\citenamefont {Badurina}\ \emph {et~al.}(2025)\citenamefont {Badurina}, \citenamefont {Du}, \citenamefont {Lee}, \citenamefont {Wang},\ and\ \citenamefont {Zurek}}]{Badurina2025}%
  \BibitemOpen
  \bibfield  {author} {\bibinfo {author} {\bibfnamefont {L.}~\bibnamefont {Badurina}}, \bibinfo {author} {\bibfnamefont {Y.}~\bibnamefont {Du}}, \bibinfo {author} {\bibfnamefont {V.~S.}\ \bibnamefont {Lee}}, \bibinfo {author} {\bibfnamefont {Y.}~\bibnamefont {Wang}},\ and\ \bibinfo {author} {\bibfnamefont {K.~M.}\ \bibnamefont {Zurek}},\ }\bibfield  {title} {\bibinfo {title} {Signatures of linearized gravity in atom interferometers: A simplified computational framework},\ }\href {https://doi.org/10.1103/PhysRevD.111.042002} {\bibfield  {journal} {\bibinfo  {journal} {Phys. Rev. D}\ }\textbf {\bibinfo {volume} {111}},\ \bibinfo {pages} {042002} (\bibinfo {year} {2025})}\BibitemShut {NoStop}%
\bibitem [{\citenamefont {Roura}(2025)}]{Roura2025}%
  \BibitemOpen
  \bibfield  {author} {\bibinfo {author} {\bibfnamefont {A.}~\bibnamefont {Roura}},\ }\bibfield  {title} {\bibinfo {title} {Atom interferometer as a freely falling clock for time-dilation measurements},\ }\href {https://doi.org/10.1088/2058-9565/ad9e2e} {\bibfield  {journal} {\bibinfo  {journal} {Quantum Science and Technology}\ }\textbf {\bibinfo {volume} {10}},\ \bibinfo {pages} {025004} (\bibinfo {year} {2025})}\BibitemShut {NoStop}%
\bibitem [{\citenamefont {Niehof}\ \emph {et~al.}(2025)\citenamefont {Niehof}, \citenamefont {Derr},\ and\ \citenamefont {Giese}}]{niehof2025finite}%
  \BibitemOpen
  \bibfield  {author} {\bibinfo {author} {\bibfnamefont {C.}~\bibnamefont {Niehof}}, \bibinfo {author} {\bibfnamefont {D.}~\bibnamefont {Derr}},\ and\ \bibinfo {author} {\bibfnamefont {E.}~\bibnamefont {Giese}},\ }\bibfield  {title} {\bibinfo {title} {Finite-speed-of-light effects in atom interferometry: Diffraction mechanisms and resonance conditions},\ }\bibfield  {journal} {\bibinfo  {journal} {AVS Quantum Science}\ }\textbf {\bibinfo {volume} {7}},\ \href {https://doi.org/10.1116/5.0284803} {10.1116/5.0284803} (\bibinfo {year} {2025})\BibitemShut {NoStop}%
\bibitem [{\citenamefont {Overstreet}\ \emph {et~al.}(2021)\citenamefont {Overstreet}, \citenamefont {Asenbaum},\ and\ \citenamefont {Kasevich}}]{overstreet2021physically}%
  \BibitemOpen
  \bibfield  {author} {\bibinfo {author} {\bibfnamefont {C.}~\bibnamefont {Overstreet}}, \bibinfo {author} {\bibfnamefont {P.}~\bibnamefont {Asenbaum}},\ and\ \bibinfo {author} {\bibfnamefont {M.~A.}\ \bibnamefont {Kasevich}},\ }\bibfield  {title} {\bibinfo {title} {Physically significant phase shifts in matter-wave interferometry},\ }\href {https://doi.org/10.1119/10.0002638} {\bibfield  {journal} {\bibinfo  {journal} {Am. J. Phys.}\ }\textbf {\bibinfo {volume} {89}},\ \bibinfo {pages} {324} (\bibinfo {year} {2021})}\BibitemShut {NoStop}%
\bibitem [{\citenamefont {Antoine}(2006)}]{antoine2006matter}%
  \BibitemOpen
  \bibfield  {author} {\bibinfo {author} {\bibfnamefont {C.}~\bibnamefont {Antoine}},\ }\bibfield  {title} {\bibinfo {title} {Matter wave beam splitters in gravito-inertial and trapping potentials: generalized ttt scheme for atom interferometry},\ }\href {https://doi.org/10.1007/s00340-006-2378-8} {\bibfield  {journal} {\bibinfo  {journal} {Applied Physics B}\ }\textbf {\bibinfo {volume} {84}},\ \bibinfo {pages} {585} (\bibinfo {year} {2006})}\BibitemShut {NoStop}%
\bibitem [{\citenamefont {Jansen}(2007)}]{jansen2007atom}%
  \BibitemOpen
  \bibfield  {author} {\bibinfo {author} {\bibfnamefont {M.~A. H.~M.}\ \bibnamefont {Jansen}},\ }\emph {\bibinfo {title} {Atom interferometry with cold metastable helium}},\ \href {https://doi.org/10.6100/IR623089} {Ph.D. thesis},\ \bibinfo  {school} {Technische Universiteit Eindhoven} (\bibinfo {year} {2007})\BibitemShut {NoStop}%
\bibitem [{\citenamefont {Glick}\ and\ \citenamefont {Kovachy}(2026)}]{glick2026feynman}%
  \BibitemOpen
  \bibfield  {author} {\bibinfo {author} {\bibfnamefont {J.}~\bibnamefont {Glick}}\ and\ \bibinfo {author} {\bibfnamefont {T.}~\bibnamefont {Kovachy}},\ }\bibfield  {title} {\bibinfo {title} {Feynman diagrams for matter wave interferometry},\ }\bibfield  {journal} {\bibinfo  {journal} {AVS Quantum Science}\ }\textbf {\bibinfo {volume} {8}},\ \href {https://doi.org/10.1116/5.0313350} {10.1116/5.0313350} (\bibinfo {year} {2026})\BibitemShut {NoStop}%
\bibitem [{\citenamefont {Le~Gou{\"e}t}\ \emph {et~al.}(2007)\citenamefont {Le~Gou{\"e}t}, \citenamefont {Cheinet}, \citenamefont {Kim}, \citenamefont {Holleville}, \citenamefont {Clairon}, \citenamefont {Landragin},\ and\ \citenamefont {Pereira Dos~Santos}}]{le2007influence}%
  \BibitemOpen
  \bibfield  {author} {\bibinfo {author} {\bibfnamefont {J.}~\bibnamefont {Le~Gou{\"e}t}}, \bibinfo {author} {\bibfnamefont {P.}~\bibnamefont {Cheinet}}, \bibinfo {author} {\bibfnamefont {J.}~\bibnamefont {Kim}}, \bibinfo {author} {\bibfnamefont {D.}~\bibnamefont {Holleville}}, \bibinfo {author} {\bibfnamefont {A.}~\bibnamefont {Clairon}}, \bibinfo {author} {\bibfnamefont {A.}~\bibnamefont {Landragin}},\ and\ \bibinfo {author} {\bibfnamefont {F.}~\bibnamefont {Pereira Dos~Santos}},\ }\bibfield  {title} {\bibinfo {title} {Influence of lasers propagation delay on the sensitivity of atom interferometers},\ }\href {https://doi.org/10.1140/epjd/e2007-00218-2} {\bibfield  {journal} {\bibinfo  {journal} {The European Physical Journal D}\ }\textbf {\bibinfo {volume} {44}},\ \bibinfo {pages} {419} (\bibinfo {year} {2007})}\BibitemShut {NoStop}%
\bibitem [{\citenamefont {Cheng}\ \emph {et~al.}(2015)\citenamefont {Cheng}, \citenamefont {Gillot}, \citenamefont {Merlet},\ and\ \citenamefont {Pereira Dos~Santos}}]{cheng2015influence}%
  \BibitemOpen
  \bibfield  {author} {\bibinfo {author} {\bibfnamefont {B.}~\bibnamefont {Cheng}}, \bibinfo {author} {\bibfnamefont {P.}~\bibnamefont {Gillot}}, \bibinfo {author} {\bibfnamefont {S.}~\bibnamefont {Merlet}},\ and\ \bibinfo {author} {\bibfnamefont {F.}~\bibnamefont {Pereira Dos~Santos}},\ }\bibfield  {title} {\bibinfo {title} {Influence of chirping the raman lasers in an atom gravimeter: Phase shifts due to the raman light shift and to the finite speed of light},\ }\href {https://doi.org/10.1103/PhysRevA.92.063617} {\bibfield  {journal} {\bibinfo  {journal} {Physical Review A}\ }\textbf {\bibinfo {volume} {92}},\ \bibinfo {pages} {063617} (\bibinfo {year} {2015})}\BibitemShut {NoStop}%
\bibitem [{\citenamefont {Swan}\ and\ \citenamefont {Hogan}(2025)}]{swan2025atom}%
  \BibitemOpen
  \bibfield  {author} {\bibinfo {author} {\bibfnamefont {H.}~\bibnamefont {Swan}}\ and\ \bibinfo {author} {\bibfnamefont {J.~M.}\ \bibnamefont {Hogan}},\ }\bibfield  {title} {\bibinfo {title} {Atom interferometer phase shear and spacetime sectional curvature},\ }\bibfield  {journal} {\bibinfo  {journal} {arXiv preprint arXiv:2508.21331}\ }\href {https://doi.org/10.48550/arXiv.2508.21331} {10.48550/arXiv.2508.21331} (\bibinfo {year} {2025})\BibitemShut {NoStop}%
\bibitem [{\citenamefont {Graham}\ \emph {et~al.}(2016)\citenamefont {Graham}, \citenamefont {Hogan}, \citenamefont {Kasevich},\ and\ \citenamefont {Rajendran}}]{graham2016resonant}%
  \BibitemOpen
  \bibfield  {author} {\bibinfo {author} {\bibfnamefont {P.~W.}\ \bibnamefont {Graham}}, \bibinfo {author} {\bibfnamefont {J.~M.}\ \bibnamefont {Hogan}}, \bibinfo {author} {\bibfnamefont {M.~A.}\ \bibnamefont {Kasevich}},\ and\ \bibinfo {author} {\bibfnamefont {S.}~\bibnamefont {Rajendran}},\ }\bibfield  {title} {\bibinfo {title} {Resonant mode for gravitational wave detectors based on atom interferometry},\ }\href {https://doi.org/10.1103/PhysRevD.94.104022} {\bibfield  {journal} {\bibinfo  {journal} {Phys. Rev. D}\ }\textbf {\bibinfo {volume} {94}},\ \bibinfo {pages} {104022} (\bibinfo {year} {2016})}\BibitemShut {NoStop}%
\bibitem [{\citenamefont {Born}(1909{\natexlab{a}})}]{born1909theorie}%
  \BibitemOpen
  \bibfield  {author} {\bibinfo {author} {\bibfnamefont {M.}~\bibnamefont {Born}},\ }\bibfield  {title} {\bibinfo {title} {Die theorie des starren elektrons in der kinematik des relativit{\"a}tsprinzips},\ }\href@noop {} {\bibfield  {journal} {\bibinfo  {journal} {Annalen der Physik}\ }\textbf {\bibinfo {volume} {335}},\ \bibinfo {pages} {1} (\bibinfo {year} {1909}{\natexlab{a}})}\BibitemShut {NoStop}%
\bibitem [{\citenamefont {Born}(1909{\natexlab{b}})}]{born1909dynamik}%
  \BibitemOpen
  \bibfield  {author} {\bibinfo {author} {\bibfnamefont {M.}~\bibnamefont {Born}},\ }\bibfield  {title} {\bibinfo {title} {{\"U}ber die dynamik des elektrons in der kinematik des relativit{\"a}tsprinzips},\ }\href@noop {} {\bibfield  {journal} {\bibinfo  {journal} {Physikalische Zeitschrift}\ }\textbf {\bibinfo {volume} {10}},\ \bibinfo {pages} {814} (\bibinfo {year} {1909}{\natexlab{b}})}\BibitemShut {NoStop}%
\bibitem [{\citenamefont {Swan}\ and\ \citenamefont {Hogan}(2026)}]{math_supp}%
  \BibitemOpen
  \bibfield  {author} {\bibinfo {author} {\bibfnamefont {H.}~\bibnamefont {Swan}}\ and\ \bibinfo {author} {\bibfnamefont {J.~M.}\ \bibnamefont {Hogan}},\ }\href@noop {} {\bibinfo {title} {exact-atom-phase}},\ \bibinfo {howpublished} {\url{https://github.com/hoganphysics/exact-atom-phase}} (\bibinfo {year} {2026})\BibitemShut {NoStop}%
\bibitem [{\citenamefont {Jiang}\ \emph {et~al.}(2025)\citenamefont {Jiang}, \citenamefont {Rudolph},\ and\ \citenamefont {Hogan}}]{jiang2025cumulative}%
  \BibitemOpen
  \bibfield  {author} {\bibinfo {author} {\bibfnamefont {Y.}~\bibnamefont {Jiang}}, \bibinfo {author} {\bibfnamefont {J.}~\bibnamefont {Rudolph}},\ and\ \bibinfo {author} {\bibfnamefont {J.~M.}\ \bibnamefont {Hogan}},\ }\bibfield  {title} {\bibinfo {title} {Cumulative fidelity of lmt clock atom interferometers in the presence of laser noise},\ }\bibfield  {journal} {\bibinfo  {journal} {arXiv preprint arXiv:2508.10288}\ }\href {https://doi.org/10.48550/arXiv.2508.10288} {10.48550/arXiv.2508.10288} (\bibinfo {year} {2025})\BibitemShut {NoStop}%
\bibitem [{\citenamefont {Dubetsky}\ and\ \citenamefont {Kasevich}(2006)}]{dubetsky2006atom}%
  \BibitemOpen
  \bibfield  {author} {\bibinfo {author} {\bibfnamefont {B.}~\bibnamefont {Dubetsky}}\ and\ \bibinfo {author} {\bibfnamefont {M.}~\bibnamefont {Kasevich}},\ }\bibfield  {title} {\bibinfo {title} {Atom interferometer as a selective sensor of rotation or gravity},\ }\href {https://doi.org/10.1103/PhysRevA.74.023615} {\bibfield  {journal} {\bibinfo  {journal} {Physical Review A}\ }\textbf {\bibinfo {volume} {74}},\ \bibinfo {pages} {023615} (\bibinfo {year} {2006})}\BibitemShut {NoStop}%
\bibitem [{\citenamefont {Lan}\ \emph {et~al.}(2012)\citenamefont {Lan}, \citenamefont {Kuan}, \citenamefont {Estey}, \citenamefont {Haslinger},\ and\ \citenamefont {M{\"u}ller}}]{lan2012influence}%
  \BibitemOpen
  \bibfield  {author} {\bibinfo {author} {\bibfnamefont {S.-Y.}\ \bibnamefont {Lan}}, \bibinfo {author} {\bibfnamefont {P.-C.}\ \bibnamefont {Kuan}}, \bibinfo {author} {\bibfnamefont {B.}~\bibnamefont {Estey}}, \bibinfo {author} {\bibfnamefont {P.}~\bibnamefont {Haslinger}},\ and\ \bibinfo {author} {\bibfnamefont {H.}~\bibnamefont {M{\"u}ller}},\ }\bibfield  {title} {\bibinfo {title} {Influence of the coriolis force in atom interferometry},\ }\href {https://doi.org/10.1103/PhysRevLett.108.090402} {\bibfield  {journal} {\bibinfo  {journal} {Physical Review Letters}\ }\textbf {\bibinfo {volume} {108}},\ \bibinfo {pages} {090402} (\bibinfo {year} {2012})}\BibitemShut {NoStop}%
\end{thebibliography}%

\end{document}